\documentclass[preprint,12pt]{elsarticle}

\usepackage{amssymb}

\usepackage{amsthm}
\usepackage[usenames,dvipsnames]{pstricks}
\usepackage{epsfig}
\usepackage{pst-grad} 
\usepackage{pst-plot} 
\usepackage{graphicx,subfigure}
\usepackage{amsmath,amssymb}

\usepackage{acronym}
\acrodef{BO}{{\bf Benjamin-Ono}}
\acrodef{rBO}{{\bf regularized Benjamin-Ono}}
\acrodef{rILW}{{\bf regularized Intermediate Long Wave}}
\acrodef{DSW}{{\bf Dispersive Shock Wave}}
\acrodef{DSWs}{{\bf Dispersive Shock Waves}}
\acrodef{ILW}{{\bf Intermediate Long Wave}}
\acrodef{CGN}{{\bf Conjugate Gradient-Newton}}
\acrodef{SW/SW}{{\bf Shallow water / Shallow water}}
\acrodef{B/B}{{\bf Boussinesq / Boussinesq}}

\makeatletter
\newcommand{\sech}{\mathop{\operator@font sech}}
\newcommand{\sign}{\mathop{\operator@font sign}}
\makeatother

\newtheorem{remark}{Remark}[section]

\begin{document}

\begin{frontmatter}



\title{Solitary-wave solutions of Benjamin-Ono and other systems for internal waves. II. Dynamics}


\author[label0]{Jerry Bona}
\address[label0]{University of Illinois at Chicago, Department of Mathematics, Statistics and Computer Science, SEO 1206, 851 South Morgan Street, Chicago, IL 60607, USA}
\ead{jbona@uic.edu}

\author[label1]{Angel Dur\'an}
\address[label1]{Applied Mathematics Department, University of Valladolid, P/ Bel\'en 7, 47011, Valladolid, Spain}
\ead{angeldm@uva.es}

\author[label2]{Dimitrios Mitsotakis}
\address[label2]{Victoria University of Wellington, School of Mathematics and Statistics, PO Box 600, Wellington 6140, New Zealand}
\ead{dimitrios.mitsotakis@vuw.ac.nz}
\ead[url]{http://dmitsot.googlepages.com/}

\begin{abstract}
Considered here are two systems of equations modeling the two-way propagation of long-crested, long-wavelength internal waves along the interface of a two-layer system of fluids in the Benjamin-Ono and the Intermediate Long-Wave regime, respectively. These systems were previously shown to have solitary-wave solutions, decaying to zero algebraically for the Benjamin-Ono system, and exponentially in the Intermediate Long-Wave regime. Several methods to approximate solitary-wave profiles were introduced and analyzed by the authors in Part I of this project. A natural continuation of this previous work, pursued here, is to study the dynamics of the solitary-wave solutions of these systems. This will be done by computational means using a  discretization of the periodic initial-value problem. The numerical method used here is a Fourier spectral 
method for the spatial approximation coupled with a fourth-order, explicit Runge-Kutta time stepping.  The resulting, fully discrete scheme is used to study computationally the stability of the solitary waves under small and large perturbations, the collisions of solitary waves, the resolution of initial data into  trains of solitary waves, and the formation of dispersive shock waves. Comparisons with related unidirectional models are also undertaken.
\end{abstract}

\begin{keyword}
Internal waves \sep Benjamin-Ono and Intermediate Long Wave systems \sep solitary
waves  \sep pseudospectral methods


\MSC 76B25 \sep 35Q35 \sep 65M70

\end{keyword}

\end{frontmatter}



\section{Introduction}\label{sec:intro}
In the precursor \cite{BDM2021} to this work, the authors considered Benjamin-Ono and  Intermediate Long Wave  systems 
of the form
\begin{equation}\label{E1}
\begin{array}{l}
\left[1+\sqrt{\mu}\frac{\alpha}{\gamma}\mathcal{H} \right]\zeta_t+\frac{1}{\gamma}\left((1-\varepsilon \zeta)u \right)_x-(1-\alpha)\frac{\sqrt{\mu}}{\gamma^2}\mathcal{H}u_x=0,\\
u_t+(1-\gamma)\zeta_x-\frac{\varepsilon}{2\gamma}(u^2)_x=0,\ 
\end{array}
\end{equation}
for $x \in \mathbb R$ and $t \geq 0$.  These were derived in \cite{BLS} and are here written in unscaled, dimensionless variables.   
 The system (\ref{E1}) is a one-spatial-dimensional model for the two-way propagation of long-crested internal waves along the interface of a two-layer system of homogeneous, inviscid fluids of densities $\rho_{j}, j=1,2$, with the bottom layer density  $\rho_{2}>\rho_{1}$ for static stability, and depths $d_{j}, j=1,2$. The upper layer is taken to be bounded above by a rigid surface (the so-called {\it rigid lid approximation}) and the lower layer is bounded below by a horizontal, flat, impermeable bottom. 
 
 As mentioned, these models allow for bi-directional propagation, unlike the original Benjamin-Ono and Intermediate Long-Wave equations (abbreviated to BO and ILW henceforth).  However, they do assume the waves are long-crested, so there is no appreciable variation in the horizontal direction orthogonal to the principle direction of propagation.   The independent  variable $x$ is proportional to position in the direction of propagation while $t$ is proportional to elapsed time, $\zeta=\zeta(x,t)$ is the deviation of the interface from its  rest position at the point $x$ at time $t$.  The dependent variable $u=u(x,t)$ is a horizontal velocity-like variable while 
  $\gamma=\rho_{1}/\rho_{2}<1$ is the density ratio and  $\alpha$ is a modeling parameter.

In terms of the non-dimensional parameters
\begin{equation*}
\epsilon=\frac{a}{d_{1}}, \;\; \mu=\frac{d_{1}^{2}}{\lambda^{2}}, \;\; \epsilon_{2}=\frac{a}{d_{2}}=\epsilon\delta \;\; {\rm and}  \;\; \mu_{2}=\frac{d_{2}^{2}}{\lambda^{2}}=\frac{\mu}{\delta^{2}},
\end{equation*}
(where $a$ and $\lambda$ denote, respectively, a typical amplitude and wavelength of the interfacial wave and                   $\delta=d_{1}/d_{2}$ is the depth ratio), the physical regimes under which the Euler equations for internal waves are consistent with the two-dimensional version of (\ref{E1}) are (see \cite{BLS}):
\begin{itemize}
\item Intermediate Long-Wave regime: $\mu\sim\epsilon^{2}\sim\epsilon_{2}\ll1, \, \mu_{2}\sim 1$. (This means that the upper layer is shallow, and the interfacial deformations are small with respect to the depths of both layers.)
\item Benjamin-Ono regime: $\mu\sim\epsilon^{2}\ll1, \, \epsilon_{2}\ll1,\,  \mu_{2}=+\infty$ (corresponding to
a very deep lower layer).
\end{itemize}
The two regimes are mathematically distinguished in (\ref{E1}) by the definition of the Fourier multiplier operator $\mathcal{H}$. This takes the form $\mathcal{H}=\partial_{x}\mathbb{T}_{\sqrt{\mu_{2}}}$, with
\begin{equation}\label{eq:PVILW}
\mathbb{T}_{\beta}\zeta=\frac{1}{2\beta}   \mbox{P.V.}\int_{-\infty}^\infty \coth\frac{\pi(\xi-x)}{2\beta}\zeta(\xi,t)\ d\xi,\; \;\;\beta>0,
\end{equation}
in the \acs{ILW} case (where P.V. stands for the Cauchy principal value of the integral), and $\mathcal{H}=\partial_{x}\mathbb{H}$, where
\begin{equation}\label{eq:PV}
\mathbb{H}\zeta=\frac{1}{\pi}   \mbox{P.V.}\int_{-\infty}^\infty \frac{\zeta(\xi,t)}{\xi-x}\ d\xi,
\end{equation}
is the Hilbert transform, in the \acs{BO} case. In terms of the corresponding Fourier symbols, if $\widehat{f}$ denotes the Fourier transform of $f$, then $\widehat{\mathcal{H}f}(k)=g(k)\widehat{f}(k), k\in\mathbb{R}$, where
\begin{equation}\label{fsymbol}
g(k)=\left\{\begin{matrix}
|k|&\mbox{in the \acs{BO} case,}\\
k\coth(\sqrt{\mu_{2}}k)&\mbox{in the \acs{ILW} case.}
\end{matrix}\right.
\end{equation}

Briefly described now are  some mathematical properties of the initial-value problem for (\ref{E1}) (see \cite{BDM2021} for details).
The system (\ref{E1}) is linearly well posed if and only if $\alpha\geq 1$ \cite{BLS} and the nonlinear initial-value 
problem for the  \acs{ILW} case has been shown in \cite{Xu2012} to be well posed, locally in time, in suitable Sobolev spaces if $\alpha>1$. In the same paper, again when $\alpha > 1$, similar well-posedness results for the \acs{BO} system are suggested to hold because of the convergence of solutions of the \acs{ILW} system to those of the \acs{BO} system 
as $\mu_2 \to \infty$.

 On the other hand, to the best of our knowledge, the system (\ref{E1}) for $\alpha\neq 0$ has only the simple conservation laws 
\begin{equation}\label{cquants}
I_{1}(\zeta,u)=\int_{-\infty}^{\infty}\zeta~ dx,\quad I_{2}(\zeta,u)=\int_{-\infty}^{\infty}u~ dx,
\end{equation}
whereas the case $\alpha=0$, which is ill posed \cite{CGK2005} admits two additional invariant quantities, and  a Hamiltonian structure emerges.  

Another important property of (\ref{E1}) is the existence of solitary-wave solutions of  the form $\zeta(x,t)=\zeta(x-c_{s}t), \,u(x,t)=u(x-c_{s}t)$, moving with constant speed $c_{s}\neq 0$, having profiles $\zeta=\zeta(X), \,u=u(X)$ which, along with their derivatives, decay to zero as $X=x-c_{s}t\rightarrow \pm\infty$. They must be solutions of the system
\begin{equation}\label{E6}
\begin{array}{l}
\begin{pmatrix}-c_s(1+\frac{\sqrt{\mu}\alpha}{\gamma}~\mathcal{H}) & \frac{1}{\gamma}+(\alpha-1)\frac{\sqrt{\mu}}{\gamma^2}~\mathcal{H}\\
1-\gamma & -c_s \end{pmatrix}~ \begin{pmatrix}\zeta \\ u \end{pmatrix}=\frac{\varepsilon}{\gamma}\begin{pmatrix} \zeta u \\  u^2 /2\end{pmatrix}.
\end{array}
\end{equation}
Existence of smooth, small-amplitude solutions of (\ref{E6}) was recently proved by Angulo-Pava and Saut in \cite{AnguloPS2019}. In the same paper, the decay as $|X|\rightarrow\infty$ was shown to be exponential in the \acs{ILW} case and algebraic (as $1/|X|^{2}$) in the \acs{BO} case, just as for the solitary-wave solutions of their undirectional counterparts.

The numerical generation of solitary-wave solutions of (\ref{E1}) was discussed by the authors in \cite{BDM2021}, which is Part I of the present study (and see also \cite{DDS2021}). Three iterative methods, one of Newton type, the classical Petviashvili iteration, and a modification of it, were proposed to solve iteratively a discretization of (\ref{E6}) based on Fourier collocation approximation of the corresponding periodic problem. The three schemes were compared in accuracy and some properties of the resulting computed solitary waves that emerged from the experiments were pointed out. These concerned the speed-amplitude relationship, the convergence of the solitary waves of the \acs{ILW} system to those of the \acs{BO} system, and  comparisons with solitary-wave solutions of both the  classical and regularized versions of the unidirectional \acs{BO} and \acs{ILW}.  The regularized versions of these unidirectional equations, which result from using the lowest-order relation, $\partial_x = -\partial_t$ to modify the dispersion relation, will be denoted henceforth by \acs{rBO} and \acs{rILW}, respectively.

With a reasonably good grasp of the solitary-wave solutions in hand, it is natural to turn to their  dynamics as solutions of the time-dependent problem. To this end, the corresponding periodic initial-value problem (IVP) is discretized in space with a spectral  method and in time with the explicit, fourth-order Runge-Kutta (RK) scheme. Error estimates of the spectral semi-discretization were proved in \cite{DDS2021}. These estimates naturally depend  upon the regularity of the solutions of (\ref{E1}) and, in particular, lead to spectral convergence in the smooth case. In addition, the fully discrete method was also used in Part I of this project to check the accuracy of the computed solitary-wave profiles. Its very satisfactory  performance gave the confidence needed to make use of it for the developments in the present essay.

From this computational perspective,  several properties of (\ref{E1}) are studied. The first group of experiments analyzes the stability  of the solitary-wave solutions. Small perturbations of the traveling-wave profiles, which are determined  as in \cite{BDM2021}, are taken as initial conditions of the time-dependent numerical scheme. The evolution of the corresponding numerical approximation suggests that the solitary waves are asymptotically stable.  
Indeed, what is observed is that a perturbed initial solitary wave develops into a principal part which 
appears to converge rapidly to a new solitary wave, together with smaller waves of a purely dispersive nature.  There are  two groups of these dispersive waves, traveling in opposite directions.  

One would expect that internal solitary waves would be unstable or lead to series of solitary waves when perturbed by a large amount. Surprisingly, the experiments suggest that this is not the case for the \acs{BO} and \acs{ILW} systems. Large perturbations of their solitary waves appear to develop into a single, large solitary wave.   The amplitude of the emergent solitary wave is related to the size of the perturbation in a way to be discussed later.  The solitary wave is followed by comparatively small amplitude,  dispersive tails quite similar to those appearing when small perturbations are considered.

Another issue studied in the present paper, which appears related to their stability, is the interaction of solitary waves. Since (\ref{E1}) admits two-way propagation, both experiments of head-on and of overtaking collisions are carried out. As the interactions are expected to be inelastic, additional information provided by the experiments will concern the emerging waves and the nature of the dispersion resulting from the collisions. We complete the computational study of properties related to the stability of solitary waves by analyzing numerically the resolution of general initial data into trains of solitary waves along with dispersive tails.
 In particular,  the role of the energy and mass of the initial data,  represented in terms of its amplitude and wavelength,  is examined in the context of this resolution property.  
 
The computational study of (\ref{E1}) will be concluded with a discussion of two additional issues.  The first one concerns the relationship between the system and the corresponding unidirectional equations. More specifically, we are interested in the dynamics of (\ref{E1}) when presented with data corresponding to unidirectional propagation. To this end,  we
analyze computationally the evolution of the interface according to the system using initial data derived from solitary-wave solutions of the unidirectional model and vice-versa.  We find that the systems compare qualitatively, and to some extent quantitatively, with their unidirectional counterparts in the unidirectional regime.  
Indeed, in both cases, the numerical experiments reveal that the evolution that emerges is similar to that which obtains from small perturbations of the exact solitary-wave solution of the relevant system. 

The second point concerns the formation of dispersive shock waves (\acs{DSW} henceforth). 
When they exist, one expects such waves to involve two scales, one fast and oscillatory and a second one, slow and modulational, {\it cf}. \cite{HoeferA2009,ElH2016}. Thus, such waves would consist of a wavetrain with a local, rapidly oscillating structure while the envelope wave parameters themselves are changing on a much slower time scale. In the present paper, the formation of \acs{DSW}s associated to (\ref{E1}) is investigated by integrating numerically two problems typically related to \acs{DSW} formation: the Riemann problem and the dam break problem.


The paper is structured as follows: Section \ref{sec:ummethod} is devoted to a description and an accuracy study of the numerical method used for the computational work to follow. In Section \ref{sec:stab},  the stability of the solitary waves is analyzed via  a group of experiments concerning the evolution of small and large perturbations of the solitary-wave 
initial data. The investigation of head-on and overtaking collision of  solitary waves, as well as the resolution property is presented in Section \ref{sec:collis}. Discussion of some relationships between unidirectional models and the associated bi-directional systems is the subject of Section \ref{sec:dyn}, while Section \ref{sec:DSWs} contains the computational study of 
\acs{DSW}s. Some concluding remarks may be found in the final Section \ref{sec:concluding}.

Overall,  the dynamics for the \acs{ILW} and the \acs{BO} systems appear  qualitatively quite similar. To keep the length of the script under control, the present paper will mostly report experiments on the dynamics of the \acs{BO} system, partly because \acs{BO}  is computationally more challenging owing to the slower decay of its solitary waves.  

\section{Description of the numerical method}
\label{sec:ummethod}
In this section the fully discrete scheme for the numerical approximation of (\ref{E1}) is described. The particular method was already used in \cite{BDM2021} to check the accuracy of the computed solitary-wave profiles.  This involved  considering 
the computed solitary-wave profiles as  initial conditions and monitoring by how much the resulting solutions
 evolve away from an appropriate 
translation of the profile.

While the problems under consideration are set on the spatial domain $\mathbb R$, the computations take place in a spatially periodic setting. 
The commonplace practice of approximating IVP's on $\mathbb R$ having localized initial data via periodic problems  goes back at least to the 1960's. (see {\it e.g.} Zabusky and Kruskal \cite{zabusky} and Tappert \cite{tappert}).  Preliminary  arguments justifying such approximation for unidirectional models can be found in  \cite{bona}, but see the more recent work of 
H. Chen  \cite{hchen}, where explicit error estimates, valid over long time scales depending on the length of the period, are obtained. The same type of analysis can be applied to the bidirectional case.
%
%

Taking this point as settled, attention is focused on the periodic IVP for (\ref{E1}) on a long enough interval $[-l,l]$, with initial conditions given by two smooth periodic functions $\zeta_{0}, u_{0}$, and where the dispersion 
operator $\mathcal{H}$ takes  the form of the corresponding  periodic operator, which is to say, $\mathcal H$ is a multiplier on the Fourier coefficients by the  function whose discrete symbol is given as in \eqref{fsymbol}   ({\it cf.} \cite{BDM2021}). 

For an even integer $N\geq 1$, let $S_{N}$ be the space of trigonometric polynomials of degree at most $N$, {\it viz.}
\begin{equation*}
S_{N}={\rm span}\big\{e^{\frac{i\pi k}{l}(x+l)}: -N\leq k\leq N\big\},\; \;\; x\in [-l,l].
\end{equation*}
Let $\{x_{j}=-l+jh, \, j=0,\ldots,2N-1 \}$ be a uniform mesh of nodes on $[-l,l]$ with $h=l/N$. Let $P=P_{N}$ denotes the $L^{2}$--projection  onto $S_{N}$ while $\mathcal{I}f =\mathcal{I}_{N}f$ denotes the interpolating trigonometric polynomial to $f$ based on the nodes $x_{j}$. 

For $T>0$, the semidiscrete Fourier-Galerkin approximation is defined as the mapping $(\zeta^{N},u^{N}):[0,T]\rightarrow S_{N}\times S_{N}$ satisfying
\begin{equation}\label{eq:pseudo1}
\begin{array}{l}
\left\langle \left[1+\sqrt{\mu}\frac{\alpha}{\gamma}\mathcal{H} \right]\zeta^N_t +\left[\frac{1}{\gamma}(1-\varepsilon \zeta^N)u^N -(1-\alpha)\frac{\sqrt{\mu}}{\gamma^2}\mathcal{H}u\right]_x,\chi \right\rangle=0, \\
\left\langle u^N_t+\left[(1-\gamma)\zeta-\frac{\varepsilon}{2\gamma}u^2 \right]_x,\chi \right\rangle=0, \quad \;\; \forall\chi\in S_N,\\
\end{array} 
\end{equation}
for $t \geq 0$, with initial data $\zeta^N(x,0)=P_N\zeta_0,~ u^N(x,0)=P_Nu_0$. The inner product $\langle\cdot,\cdot\rangle$ is that of $L^2( -l, l)$.
The spectral discretization (\ref{eq:pseudo1}) was introduced in \cite{DDS2021} and $L^{2}$-error estimates were established there. For the experiments below, and to take advantage of the numerical generation of solitary waves performed in \cite{BDM2021}, the method was implemented in pseudospectral form. Algebraically, this is equivalent to the collocation formulation \cite{CHQZ}. With a small abuse of notation, $(\zeta^{N},u^{N}):[0,T]\rightarrow S_{N}\times S_{N}$ will also denote the
semidiscrete Fourier collocation approximation, which  is to say, the mapping $(\zeta^{N},u^{N})$ satisfies
\begin{equation}\label{Enum}
\begin{array}{l}
\left[1+\sqrt{\mu}\frac{\alpha}{\gamma}\mathcal{H} \right]\zeta_t^{N}+\frac{1}{\gamma}\left(\mathcal{I}_{N}\left((1-\varepsilon \zeta^{N})u^{N}\right) \right)_x-(1-\alpha)\frac{\sqrt{\mu}}{\gamma^2}\mathcal{H}u^{N}_x=0,\\
u^{N}_t+(1-\gamma)\zeta^{N}_x-\frac{\varepsilon}{2\gamma}(\mathcal{I}_{N}(u^2))_x=0,
\end{array}
\end{equation}
at $x=x_{j}, \, j=0, \ldots, 2N-1$, with $\zeta^{N}(0)=\mathcal{I}_{N}\zeta_{0},\, u^{N}(0)=\mathcal{I}_{N}u_{0}$, so that 
\begin{equation}\label{Enum2}
\zeta^{N}(0)\big|_{x=x_{j}}=\zeta_{0}(x_{j}),\;\; u^{N}(0)\big|_{x=x_{j}}=u_{0}(x_{j}),\;\; \;0\leq j\leq 2N-1.
\end{equation}
%

The  semidiscrete system (\ref{Enum})-(\ref{Enum2}) is conveniently formulated using a nodal representation $(\zeta_{h},u_{h})$ of $(\zeta^{N},u^{N})$ with
\begin{equation*}
\begin{array}{l}
\zeta_{h}(t)=(\zeta_{h,j})_{j=0}^{2N-1},\; \;\; \zeta_{h,j}=\zeta^{N}(x_{j},t),\;\;\; j=0,\cdots,2N-1,\\
u_{h}(t)=(u_{h,j})_{j=0}^{2N-1},\;\;\;  u_{h,j}=u^{N}(x_{j},t),\; \; j=0,\cdots,2N-1.
\end{array}
\end{equation*}
The resulting system for $(\zeta_{h},u_{h})$ is 
\begin{equation}\label{Enum3}
\begin{array}{l}
\left[1+\sqrt{\mu}\frac{\alpha}{\gamma}\mathcal{H}_{N} \right]\frac{d}{dt}\zeta_{h}+\frac{1}{\gamma}\mathcal{D}_{N}\left((1-\varepsilon \zeta_{h}) \circ u_{h}\right)
 -(1-\alpha)\frac{\sqrt{\mu}}{\gamma^2}\mathcal{H}_{N}\mathcal{D}_{N}u_{h}=0,\\
\frac{d}{dt}u_{h}+(1-\gamma)\mathcal{D}_{N}\zeta_{h}-\frac{\varepsilon}{2\gamma}\mathcal{D}_{N}(u_{h}\circ u_h)=0,
\end{array}
\end{equation}
where $\circ$'s connote  Hadamard products, $\mathcal{D}_{N}$ is the Fourier pseudospectral differentiation matrix of order $2N$ and $\mathcal{H}_{N}$ stands for the corresponding discrete version of the operator $\mathcal{H}$.  In terms of the $k^{th}$ discrete Fourier coefficients $\widehat{\zeta^{N}}(k)$ and $\widehat{u^{N}}(k)$ of $\zeta^{N}$ and $u^{N}$,
the discrete Fourier transform allows us to write  (\ref{Enum3}) in the form
\begin{equation}\label{Enum4}
\begin{array}{r}
\left[1+\sqrt{\mu}\frac{\alpha}{\gamma}g_{k} \right]\frac{d}{dt}\widehat{\zeta^{N}}(k,t)+\frac{i\widetilde{k}}{\gamma}\left(\widehat{u^N} - \varepsilon \widehat{\zeta^N u^N} \right)(k,t)\\
 -(1-\alpha)\frac{\sqrt{\mu}}{\gamma^2}g_{k}(i\widetilde{k})\widehat{u^{N}}(k,t)=0,\\
\frac{d}{dt}\widehat{u^{N}}(k,t)+(1-\gamma)(i\widetilde{k})\widehat{\zeta^{N}}(k,t)-\frac{\varepsilon}{2\gamma}(i\widetilde{k})\widehat{(u^{N})^2}(k,t)=0,
\end{array}
\end{equation}
where $\widetilde{k}=\frac{\pi k}{l}
, g_{k}=g(|\widetilde{k}|), \: -N\leq k\leq N$ and $g$ is as given in (\ref{fsymbol}).

The initial-value problem for the semidiscrete system (\ref{Enum4}) is integrated in time using the explicit, $4$th-order 
Runge-Kutta method.

\begin{remark}
It is worth mentioning some additional properties of the semi-discrete schemes. Note first that for the continuous, periodic initial-value problem for the system
 (\ref{E1}) on $(-l,l)$, the analogs 
\begin{equation}\label{cquants2}
I_{1}(\zeta,u)=\int_{-l}^{l}\zeta~ dx\quad {\rm and} \quad I_{2}(\zeta,u)=\int_{-l}^{l}u~ dx, 
\end{equation}
of (\ref{cquants}) are also independent of time. A natural discretization of (\ref{cquants2}) is given by $h\mathcal{I}_{1,h}, h\mathcal{I}_{2,h}$ where
\begin{equation}\label{cquants3}
\mathcal{I}_{1,h}(\zeta_{h},u_{h})=\langle \zeta_{h},1_{N}\rangle,\quad \mathcal{I}_{2,h}(\zeta_{h},u_{h})=\langle u_{h},1_{N}\rangle,
\end{equation}
and $1_{N}$ denotes the vector with all components equal to one.  The angle brackets $\langle\cdot,\cdot\rangle$ here denote the Euclidean inner product in $\mathbb{C}^{N}$. Note that if $F_{N}$ stands for the $N\times N$ matrix associated to the discrete Fourier transform and $F_{N}^{*}$ is its adjoint, then $F_{N}^{-1}=\frac{1}{N}F_{N}^{*}$, and
\begin{equation*}
\mathcal{H}_{N}=F_{N}H_{N}F_{N}^{-1},\quad \mathcal{D}_{N}=F_{N}D_{N}F_{N}^{-1},
\end{equation*}
where $H_{N}$ (respectively $D_{N}$) denotes the diagonal matrix with diagonal entries $g_{k}$ (respectively $ik$), $k=0,\ldots,2N-1$. In particular, $H_{N}^{*}=H_{N}$ and $D_{N}^{*}=-D_{N}$. Since
\begin{equation*}
F_{N}^{-1}1_{N}=\frac{1}{N}F_{N}^{*}1_{N}=e_{1}=(1,0,\ldots,0)^{T},
\end{equation*}
then $\mathcal{H}_{N}1_{N}=\mathcal{D}_{N}1_{N}=0$. Therefore, taking the inner product with $1_{N}$ in the first and second equations of (\ref{Enum3}) reveals  that the quantities (\ref{cquants3}) are preserved up to round-off error by the semidiscrete approximation.
\end{remark}


%
%
%


\section{Stability of the solitary waves}\label{sec:stab}

\subsection{Stability of solitary waves under small and large perturbations}\label{sec:pert}
In our study of the stability of solitary waves of both \acs{ILW} and \acs{BO} systems, several types of numerical experiments are performed. The general  procedure is always the same:  first generate an approximate solitary-wave profile by using one of the methods described in \cite{BDM2021}. Then, this computed profile is perturbed in one or both  of its components $\zeta$ and $u$. Finally, the perturbed profile is used as an initial condition for the fully-discrete scheme described in Section \ref{sec:ummethod} and the evolution of the resulting numerical approximation is monitored. 

\begin{figure}[!htbp]
\centering
{\includegraphics[width=\columnwidth]{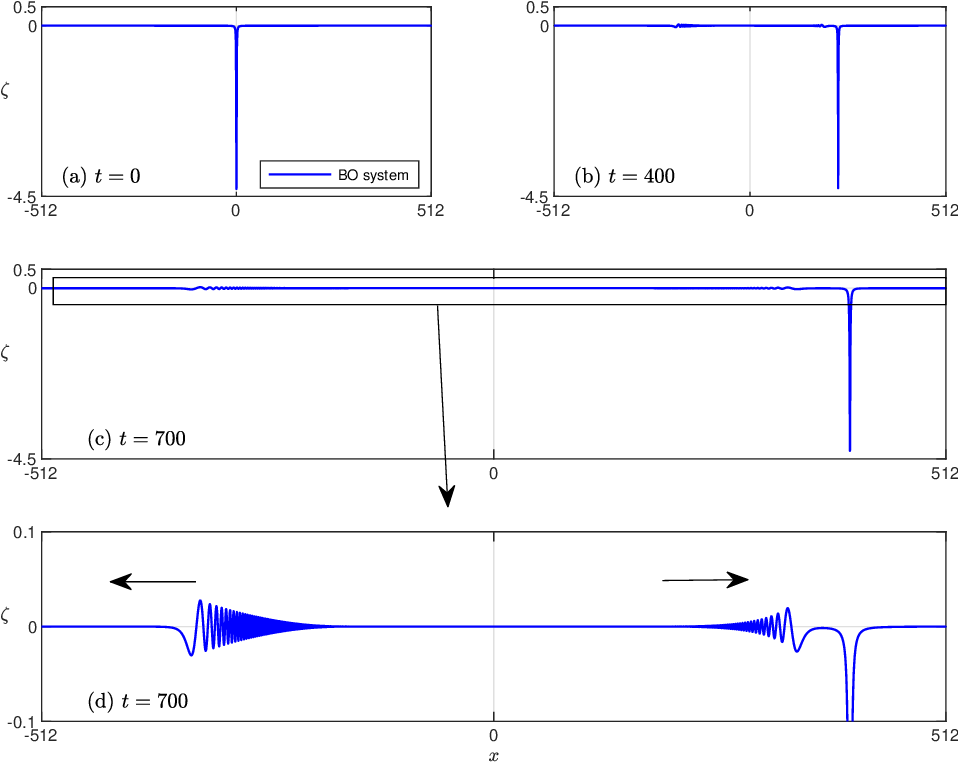}}
\caption{Nonlinear stability of a \acs{BO}-solitary wave with $c_s=0.57$ with amplitude approximately $a= 3.9129$, perturbed initially by  about $10\%$ in only the $\zeta$-component of the system. 
}
\label{fig:figure16}
\end{figure}

In all cases, the solitary waves appear to be stable under small perturbations. The behavior is illustrated by the following example: Consider  the \acs{BO} system with $\alpha=1.2$, $\gamma=0.8$ and $\varepsilon=\sqrt{\mu}=0.1$ and the solitary wave with $c_s=0.57$ and amplitude approximately $a=3.9129$.  This solitary wave is  perturbed by multiplying the function $\zeta_0(x)$ by the factor $r=1.1$. Thus the perturbed initial data is $(r\zeta_0,u_0)$ where 
$(\zeta_0,u_0)$ is the unperturbed solitary-wave profile.   This represents a perturbation of about $10\%$.  Figure \ref{fig:figure16} shows the evolution of the perturbed solitary wave to a new, stable solitary wave followed by a rather interesting dispersive tail. The dispersive tail consists of a right and a left-traveling part as indicated in Figure \ref{fig:figure16}. In this figure we only show the $\zeta$-component of the solution since $u$ behaves similarly. 
Analogous  behaviour has been observed in the case of Boussinesq systems for surface waves in \cite{DDMM}. The accuracy of the results is guaranteed in this case by using $N=32,768$ nodes in the interval $[-512,512]$ (so $h=3.125\times 10^{-2}$) and $\Delta t=0.01$. The results reported here are representative of a whole series of tests 
with $r \neq 0$, but $|r-1|$ fairly small.   

\begin{figure}[!htbp]
\centering
{\includegraphics[width=\columnwidth]{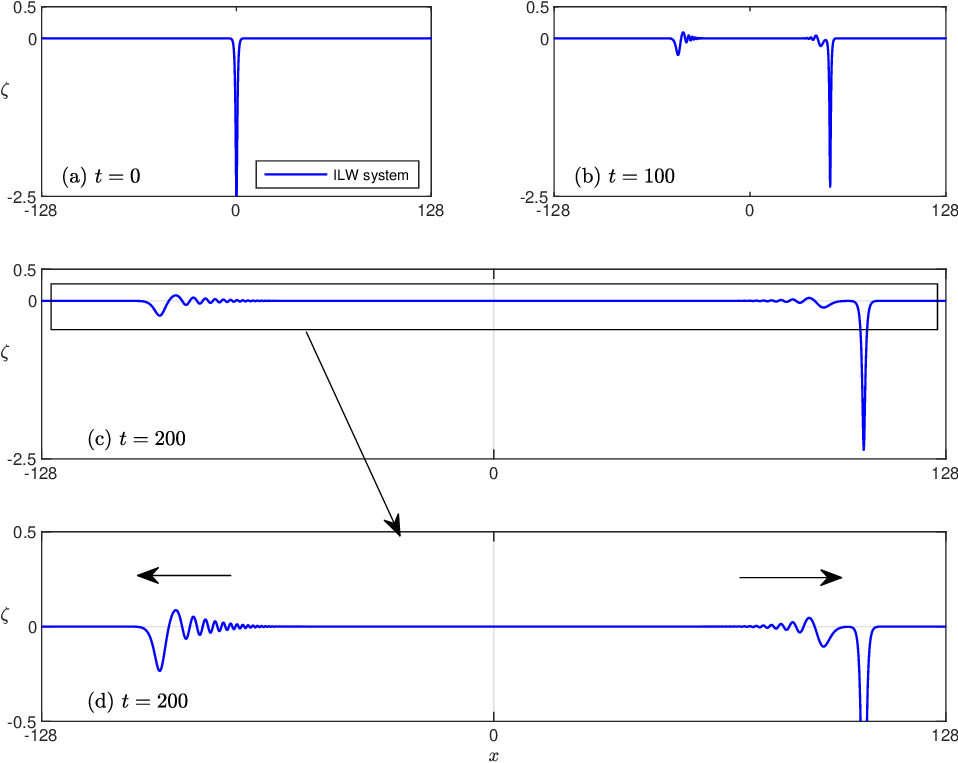}}
\caption{Nonlinear stability of a solitary wave with $c_s=0.51$ perturbed initially by  about $50\%$ in only the $\zeta$-profile of the \acs{ILW}-system. 
}
\label{fig:figure16b}
\end{figure}

The results obtained for the \acs{ILW} case are quite similar to those presented in Figure \ref{fig:figure16}.  The main  difference is that, because the solutions decays exponentially to 0,
the simulations are more easily carried out since a smaller spatial interval is required.  This stable behaviour seems to persist for larger perturbations. For example, Figure \ref{fig:figure16b} presents the evolution of a solitary wave of the \acs{ILW} system with an amplitude perturbation $r = 1.5$ in the  $\zeta$ component only. Keeping the parameters of the \acs{ILW} system the same as in the \acs{BO} system, the unperturbed speed of the solitary wave was taken to be $c_s=0.51$.  With these specifications, the amplitude was  approximately $a=1.7$. The evolution of the perturbed initial condition leads to a new solitary of amplitude approximately $a=2.35$ and two counter-propagating dispersive tails. 
A series of simulations with larger values of $r$ showed similar behavior for both \acs{BO} and \acs{ILW} systems.  
The emerging solitary wave had roughly the amplitude $rA$, where $A$ is the amplitude of the unperturbed solitary wave.
  These results speak to the extreme stability of the solitary-waves solutions of both systems.

\subsection{Dispersive tails}\label{sec:tails}
Observe in Figure \ref{fig:figure16}(d) that the tail generated by the perturbation of a solitary wave consists of two dispersive groups, a left- and a right-propagating part.  These seem to be well separated and there appears to be no interaction between them or with the emerging solitary wave.  This phenomenon has also been  observed in Boussinesq systems for surface waves.  It is worth 
remarking that in the Boussinesq system context, there were cases where an interaction between the two parts of the dispersive tails was observed; this seemed to 
occur when blowup phenomena was imminent (see  \cite{DDMM}). 

A little analysis based on observation of the outcome of 
numerical simulation casts some light on the remarks above. 
Define two quantities 
\begin{equation}\label{cgamma}
c_{\gamma}=\sqrt{(1-\gamma)/\gamma}\ 
\quad {\rm and} \quad 
R=\sqrt{\frac{\gamma+\frac{(\alpha-1)\sqrt{\mu}}{\sqrt{\mu_{2}}}}{\gamma+\frac{\alpha\sqrt{\mu}}{\sqrt{\mu_{2}}}}}\ .
\end{equation}
(Note that $R<1$.)  As seen in \cite{BDM2021}, the speed of the solitary waves seems to be bounded from below by $c_\gamma$ in case of the \acs{BO} system and by $Rc_\gamma$ in case of the \acs{ILW} systems. The solution of these systems that start from a perturbed solitary wave evolves into a solitary wave of speed $c_s$, say, 
and a small remainder, which we are calling the dispersive tail.   We suppose $c_s > c_\gamma$ in the \acs{BO} case 
and that $c_s > Rc_\gamma$ in the \acs{ILW} situation.
Small solutions of the system (\ref{E1}) satisfy approximately the linearized system
\begin{equation}\label{LE1}
\begin{array}{l}
\left[1+\sqrt{\mu}\frac{\alpha}{\gamma}\mathcal{H} \right](\partial_{t}-c_{s}\partial_{X})\zeta+\frac{1}{\gamma}\left((1- (1-\alpha)\frac{\sqrt{\mu}}{\gamma}\mathcal{H}\right)\partial_{X}u=0,\\
(\partial_{t}-c_{s}\partial_{X})u+(1-\gamma)\partial_{X}\zeta=0.
\end{array}
\end{equation}
where $X=x-c_{s}t$ is a frame of reference moving with a solitary wave of speed $c_{s}>c_{\gamma}$ in the \acs{BO} case and $c_{s}>Rc_{\gamma}$ in the \acs{ILW} case.

Since the operators $\mathcal{H}$ and $\partial_{t}-c_{s}\partial_{X}$ commute, if 
$\partial_{t}-c_{s}\partial_{X}$ is applied to the first 
equation in (\ref{LE1}) and   the second equation is used to solve for $(\partial_t -\partial_X)u$ in terms 
of $\zeta$, there obtains 
\begin{equation}\label{LE2}
\left[1+\sqrt{\mu}\frac{\alpha}{\gamma}\mathcal{H} \right](\partial_{t}-c_{s}\partial_{X})^{2}\zeta-c_{\gamma}^{2}\left(1- (1-\alpha)\frac{\sqrt{\mu}}{\gamma}\mathcal{H}\right)\partial_X^2 \zeta=0.
\end{equation}
Solutions of (\ref{LE2}) are superpositions of plane waves $\zeta(X,t)=e^{i(kX-\omega(k)t)}$ where
\begin{equation*}\label{dst1}
\omega=\omega_{\pm} (k)=-c_{s}k\pm c_{\gamma}k\phi(|k|),
\end{equation*}
and  $\phi:[0,\infty)\rightarrow \mathbb{R}$ is
\begin{equation*}\label{dst2}
\phi(x)=\sqrt{1-\frac{\frac{\sqrt{\mu}}{\gamma}g(x)}{1+\alpha\frac{\sqrt{\mu}}{\gamma}g(x)}}\, .
\end{equation*}
As before, $g$ is given in (\ref{fsymbol}). Note that $g$ is positive and increasing for $x\geq 0$. Define $g(0)=1/\sqrt{\mu_{2}}$ in the \acs{ILW} case. If one assumes $\alpha\geq 1$, which is required for well 
posedness,  then $\phi$ has the following properties:
\begin{itemize}
\item[(i)] For $x>0$,
\begin{equation}\label{dst3a}
 0<\phi(x)<\phi(0)=\left\{\begin{matrix}
1&\mbox{in the \acs{BO} case,}\\
R&\mbox{in the \acs{ILW} case.}
\end{matrix}\right. 
\end{equation}
\item[(ii)] For $x>0$,  
\begin{equation}\label{dst3}
\phi^{\prime}(x)=-\,\frac{\sqrt{\mu}}{2\gamma}\frac{g^{\prime}(x)}{\phi(x)\left(1+\alpha\frac{\sqrt{\mu}}{\gamma}g(x)\right)^{2}}<0.
\end{equation}
\item[(iii)]
$$
\lim_{x\rightarrow +\infty}\phi(x)=\sqrt{1-\frac{1}{\alpha}}\in [0,1] .
$$
\end{itemize}
Therefore, for all wavenumbers $k>0$, the phase speed
\begin{equation*}
v_{\pm} (k)=\frac{\omega_{\pm} (k)}{k}=-c_{s}\pm c_{\gamma}\phi(|k|),
\end{equation*}
of the dispersive tail, relative to the speed of the solitary wave, satisfies
\begin{equation}\label{dst4}
-(c_{s}+\phi(0)c_{\gamma})<v_{-}(k)<-c_{s}<v_{+}(k)<-c_{s}+\phi(0)c_{\gamma}.
\end{equation}
 The inequalities (\ref{dst4}) show in which direction the plane-wave components of the dispersive tails propagate, relative to the solitary wave. Since $\phi(0)<1$, the absolute phase speed $v_+(k) + c_s$ of the right-traveling waves 
 $e^{i(kX-\omega_{+}(k)t)}$ is less than or equal to $c_{\gamma}$, and similarly the left-traveling waves 
 $e^{i(kX-\omega_{-}(k)t)}$ 
 have $|v_{-}(k)+c_{s}| \leq c_{\gamma}$. Moreover, components corresponding to longer wavelength (smaller $k$) are faster than those of shorter wavelength.

 Examine now the associated group velocities. Observe that,
 \begin{equation*}\label{dst5}
 \begin{array}{l}
\omega^{\prime}_{\pm} (k)=-c_{s}\pm c_{\gamma}(\phi(|k|)+|k|\phi^{\prime}(|k|)),\quad k\neq 0,\\
\omega^{\prime}_{\pm} (0)=-c_{s}\pm c_{\gamma}\phi(0) \ .
\end{array}
\end{equation*}
For $x\geq 0$  consider the function
\begin{eqnarray*}
\psi(x)&=&\phi(x)+x\phi^{\prime}(x)=
\phi(x)-\frac{\sqrt{\mu}}{2\gamma}\frac{xg^{\prime}(x)}{\phi(x)\left(1+\alpha\frac{\sqrt{\mu}}{\gamma}g(x)\right)^{2}}\,,\quad x>0,\label{dst6} \\
\psi(0)&=&\phi(0).\nonumber
\end{eqnarray*}
Note first that (\ref{dst3}) implies that $\psi(x)\leq \phi(x)$ for $x\geq 0$. On the other hand, according to (\ref{fsymbol}),  $xg^{\prime}(x)=g(x)+x^{2}C(x)$ where
\begin{equation*}
C(x)=\left\{\begin{matrix}
0&\mbox{in the \acs{BO} case,}\\
-\frac{\sqrt{\mu_{2}}}{\sinh^{2}(\sqrt{\mu_{2}}x)}&\mbox{in the \acs{ILW} case.}
\end{matrix}\right. 
\end{equation*}
Consequently $C(x)\leq 0$ for  $x>0$ and so $xg^{\prime}(x)\geq g(x)$ for $ x> 0$. This and the hypothesis $\alpha\geq 1$ imply that
\begin{equation*}
\psi(x)\geq \phi(x)-\frac{\sqrt{\mu}}{2\gamma}\frac{g(x)}{\phi(x)\left(1+\alpha\frac{\sqrt{\mu}}{\gamma}g(x)\right)^{2}}\geq 0. 
\end{equation*}
In consequence, it transpires that 
\begin{equation*}
\begin{array}{l}
0\leq \psi(x)\leq \phi(x),\;\; {\rm for} \;\; x\geq 0, \;\; {\rm and} \\
\lim_{x\rightarrow +\infty}\psi(x)=\lim_{x\rightarrow +\infty}\phi(x)=\sqrt{1-\frac{1}{\alpha}}\ .
\end{array}
\end{equation*}
Therefore, for all wavenumbers $k>0$,
\begin{equation}\label{dst7}
-(c_{s}+\phi(0)c_{\gamma})<\omega^{\prime}_{-}(k)<-c_{s}<\omega^{\prime}_{+}(k)<-c_{s}+\phi(0)c_{\gamma}<0\ .
\end{equation}
By using (\ref{cgamma}), (\ref{dst3a}), and the hypotheses on $c_{s}$, it is deduced  that
$c_{s}>\phi(0)c_{\gamma}$ in both the \acs{BO} and \acs{ILW} cases. Then (\ref{dst7}) means that there are two dispersive groups, one traveling to the left and one to the right (with the solitary wave), but with a group velocity smaller than $c_{s}$.

\section{Solitary wave interaction and the resolution property}\label{sec:collis}

The present section is concerned with the interactions of solitary waves.  We remind the reader that for the unidirectional \acs{BO} and \acs{ILW} models, overtaking interactions 
are elastic owing to the complete integrability of the equations.  For the regularized versions of these models, apparently the collisions are not elastic as witnessed by the 
simulations in \cite{KB2000,K2005}. Our purpose now is to study computationally the interactions of solitary waves of the \acs{BO} and \acs{ILW} systems (\ref{E1}). Since the models support propagation in both directions, two types of collisions can be considered.

\subsection{Head-on collisions}\label{sec:head-on}

\begin{figure}[!htbp]
\centering
{\includegraphics[width=\columnwidth]{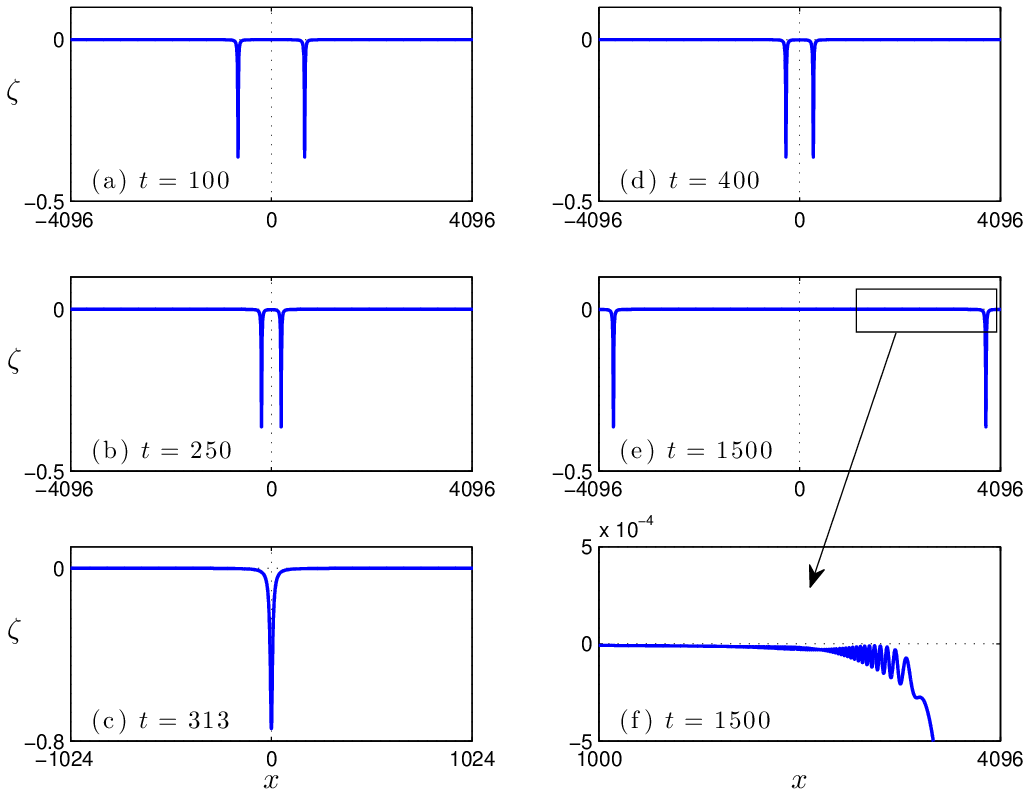}}
\caption{Symmetric head-on collision of solitary-wave solutions of the \acs{BO} system with $c_s=3.2$: (a)-(e): numerical approximation at times $t=100, 250, 313, 400, 1500$; (f) is a magnification of the marked part of (e).
}
\label{fig:figure6}
\end{figure}

The experiments begin with a {\em symmetric} head-on collision for the \acs{BO} system. This is illustrated by the following example. Two computed solitary waves with the same speed  $c_s=3.2$  (with $\alpha=1.2$, $\gamma=0.1$, $\varepsilon=0.5$, $\sqrt{\mu}=0.1$) on $[-4096,4096]$,  propagating in opposite directions are computed. (To reverse 
the direction of propagation, simply remark that if $(\zeta,u)$ is a solitary wave propagating to the right with speed $c_s$, 
then $(\zeta,-u)$ is a solitary wave propagating the the left with speed $-c_s$.). The location of the peak amplitudes of the two solitary pulses was initially placed at $x=-800$ and $x=800$, respectively. (In this and all the rest of the experiments of this section,  $h=0.125$ and $\Delta t=0.01$.) The head-on collision is illustrated in Figure \ref{fig:figure6} at several time instances of the evolution of the numerical approximation to the deviation of the interface. The interaction appears to be inelastic and similar to the head-on collision of Boussinesq systems for surface water waves, (see {\it eg.} \cite{DM2008}). During the interaction, the solitary waves change slightly in shape and their speed decreases. After the inelastic head-on collision the emerging solitary waves have altered amplitudes and speeds and they shed a symmetric dispersive tail propagating in both directions. (A magnification of the dispersive tail is shown in Figure \ref{fig:figure6}(f).)
\begin{figure}[!htbp]
\centering
{\includegraphics[width=\columnwidth]{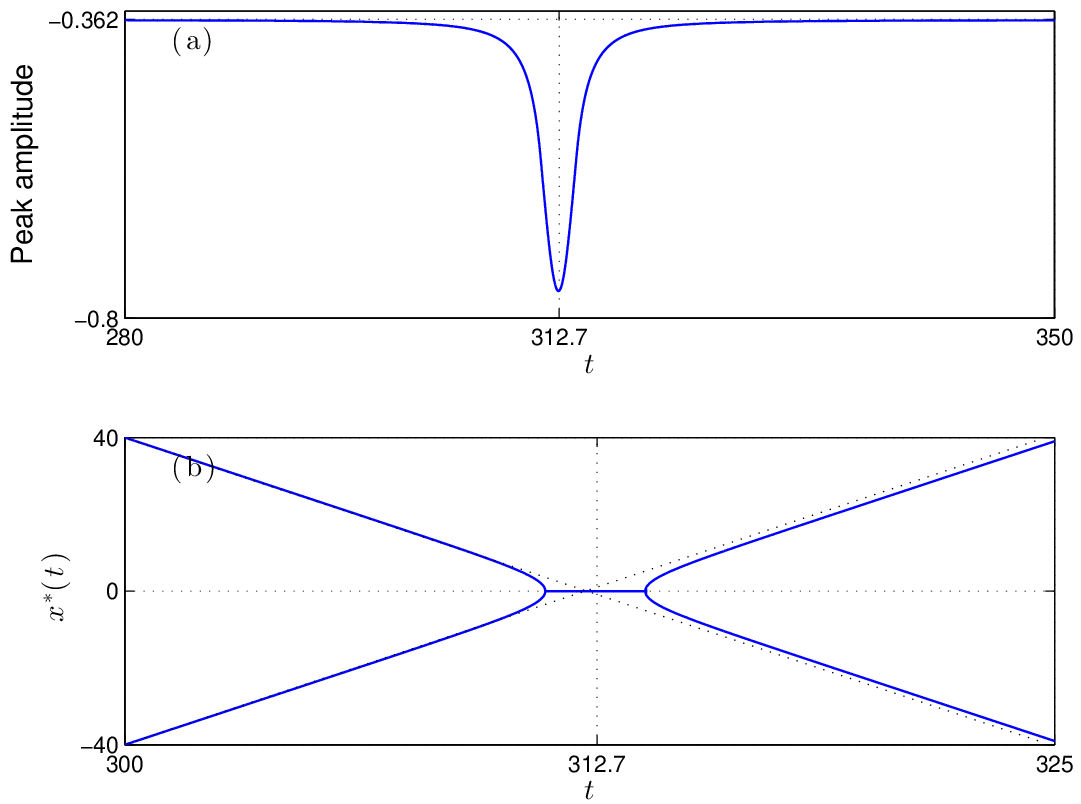}}
\caption{Location of peak amplitudes of the numerical approximation during the symmetric head-on collision  of the \acs{BO} system, see Figure \ref{fig:figure6}.  Notice the small, retarded phase shift after the interaction.
}
\label{fig:figure7}
\end{figure}
The peak amplitude of the numerical approximation during the interaction is presented in Figure \ref{fig:figure7}(a), while Figure \ref{fig:figure7}(b) shows the locations of the peak amplitudes. The amplitudes and their locations have been computed using Newton's method as described in \cite{BDM2021}. The amplitude before the collision was $A=-0.362898$ while after the (symmetric and inelastic) interaction it has decreased to $A=-0.362912$ which is a $0.39\%$ change. A 
similarly small phase change can be observed.

In addition, we performed asymmetric head-on collisions, with similar results and the same conclusions. This inelastic character is also observed in the case of the \acs{ILW} system. Due to the exponential decay rate of the solitary wave profiles 
in this case, the numerical simulations do not require such a large spatial domain for accurate approximation.

\subsection{Overtaking collisions}\label{sec:overcol}

We also consider a second type of interaction between solitary waves, namely overtaking collisions. 
These occur when a wave of one speed is placed behind another with a smaller speed.  
Recall again the elastic behavior of this type of collision for the  unidirectional \acs{ILW} and \acs{BO} equations.  
On the other hand,  
the regularized versions of these unidirectional models featured inelastic collisions (according to the experiments
reported in \cite{KB2000,K2005} for the \acs{BO} case). 

For the \acs{BO} system, the typical behavior for this interaction is illustrated by the following experiment. Taking
$\alpha=1.2$, $\gamma=0.1$ and $\varepsilon=\sqrt{\mu}=0.1$ on the interval $[-4096,4096]$, two solitary waves with speeds $c_s=3.5$ and $c_s=3.1$ (traveling to the right) with amplitudes $A_s=-4.7157 $ and $A_s=-0.8978$ respectively, have been considered as initial data for the  numerical integration of (\ref{E1}). The peaks of these two soliary waves were sited initially at $x = -1,000$ and $x=1,000$.  The numerical computations used $N=65,536$ modes (corresponding to $h=0.125$) and $\Delta t=0.01$. 

For comparison, the same experiment for the corresponding solitary-wave solutions (with the same speeds) of the \acs{rBO} model is also performed. The solitary-wave solutions for \acs{rBO} have an exact formula, and it is not necessary to generate them numerically. Having the same speeds as those for the \acs{BO} system, the amplitudes are not the same due to the different speed-amplitude relation, see \cite{BDM2021}.
%
%
\begin{figure}[!htbp]
\centering
{\includegraphics[width=\columnwidth]{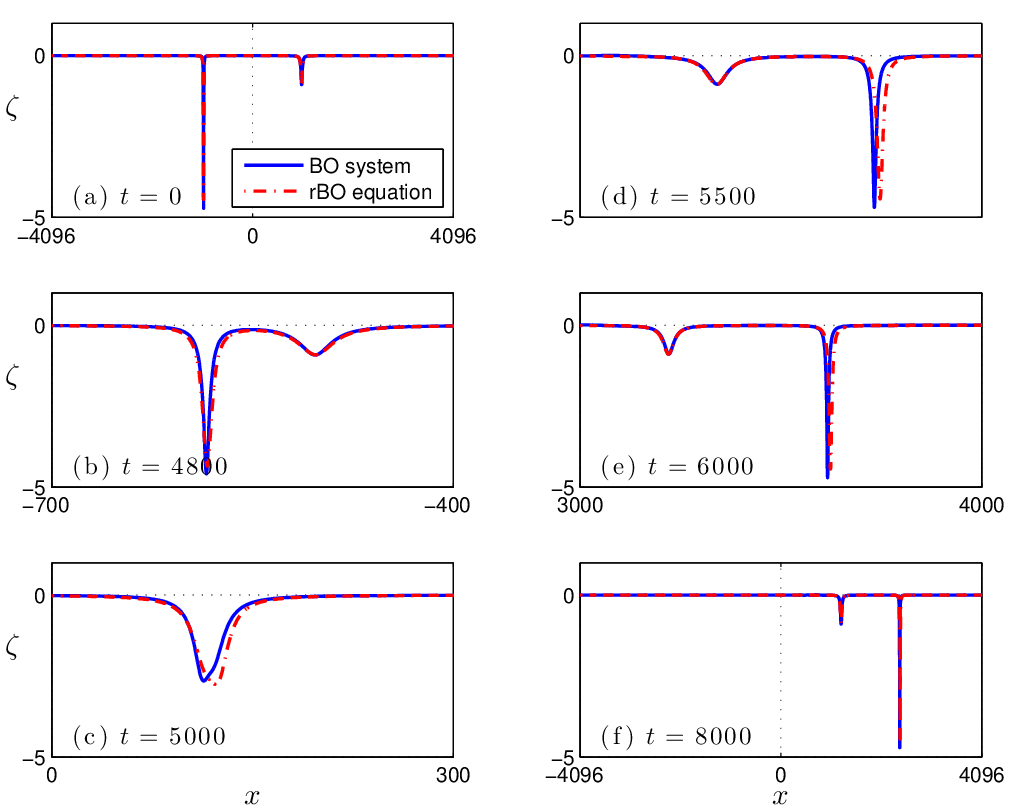}}
\caption{Overtaking collision of two solitary waves of the \acs{BO} system (solid lines) and the \acs{rBO} equation (dashed lines) with speeds $c_s=3.5$ and $c_s=3.1$.  (a)-(f): numerical approximation at times $t=0, 4800, 5000, 5500, 6000, 8000$.  After the interaction, there is a small difference in the speeds of the larger emergent solitary waves.  
}
\label{fig:figure9}
\end{figure}
Figure \ref{fig:figure9} presents the $\zeta$ profile during the interaction for different propagation times. The experiment suggests an inelastic overtaking collision of the solitary waves for both the \acs{BO} system and \acs{rBO} equation.
 We also note  (see especially \ref{fig:figure9}(d)) that the interaction of the solitary waves in the case of the \acs{rBO} equation looks more inelastic (more nonlinear) and lasts longer. Perhaps as a result of this longer interaction time, the solitary waves of the \acs{rBO} equation after the interaction appear to  possess a larger phase change. This more inelastic character in the \acs{rBO} case is noted, for example, by the fact that after the interaction, 
the larger solitary wave of the \acs{rBO} equation has a change in amplitude of about $0.68\%$, in comparison with the \acs{BO} system, where the change is  about $0.51\%$.

\begin{figure}[!htbp]
\centering
{\includegraphics[width=\columnwidth]{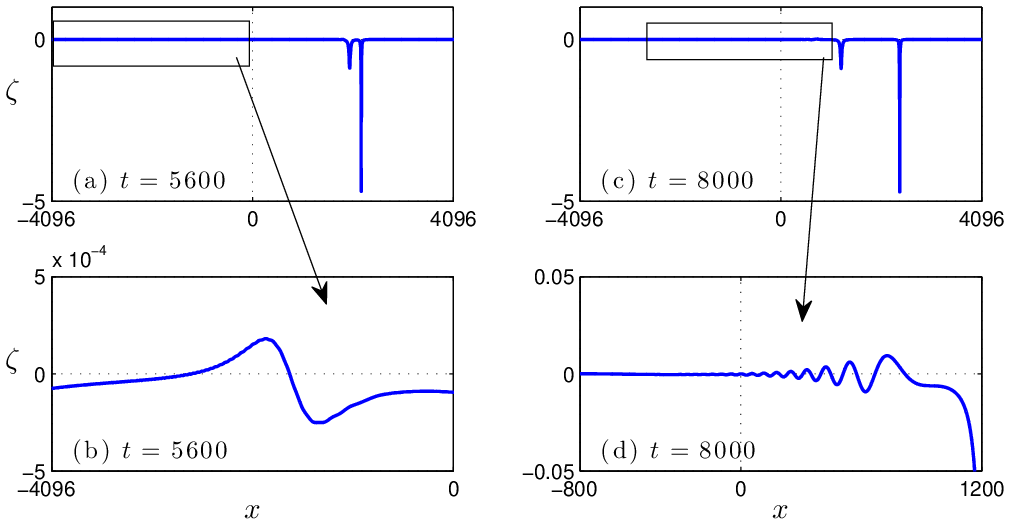}}
\caption{The dispersive tails after the interaction of the solitary waves of the \acs{BO} system appearing in Figure \ref{fig:figure9}. (b) and (d) are magnifications of the marked part of (a) and (c), respectively.   Notice the N-shaped, left-propagating wave.}
\label{fig:figure10}
\end{figure}

\begin{figure}[!htbp]
\centering
{\includegraphics[width=\columnwidth]{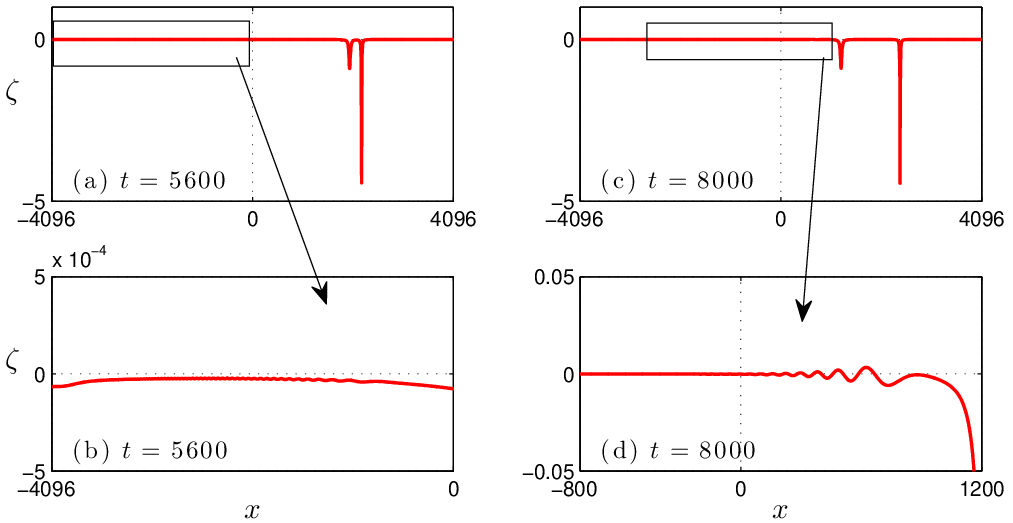}}
\caption{The dispersive tails after the interaction of the solitary waves of the \acs{rBO} equation
 from Figure \ref{fig:figure9}. Graphs (b) and (d) are magnifications of the marked part of (a) and (c), respectively. 
 There is no N-shaped residual here.}
\label{fig:figure11}
\end{figure}

As expected, after the inelastic interaction, dispersive tails were generated.
For the experiments under discussion, these emerging tails are shown in more detail in Figure \ref{fig:figure10} for the \acs{BO} system and in Figure \ref{fig:figure11} for the \acs{rBO} equation. The magnification given in Figure \ref{fig:figure10}(b)
shows that after the interaction of the solitary waves of the \acs{BO} system, an N-shaped wavelet traveling to the left is  generated. This does not appear in the case of the \acs{rBO} equation, see Figure \ref{fig:figure11}(b).
The formation of these types of waves has also been observed in unidirectional and bidirectional surface water-wave models, {\it cf.} \cite{DKM2011,DKM2012}. The rest of the dispersive tails are comparable in both size and wavelength
for the two models, see Figures \ref{fig:figure10}(d) and \ref{fig:figure11}(d).


\begin{figure}[!htbp]
\centering
{\includegraphics[width=\columnwidth]{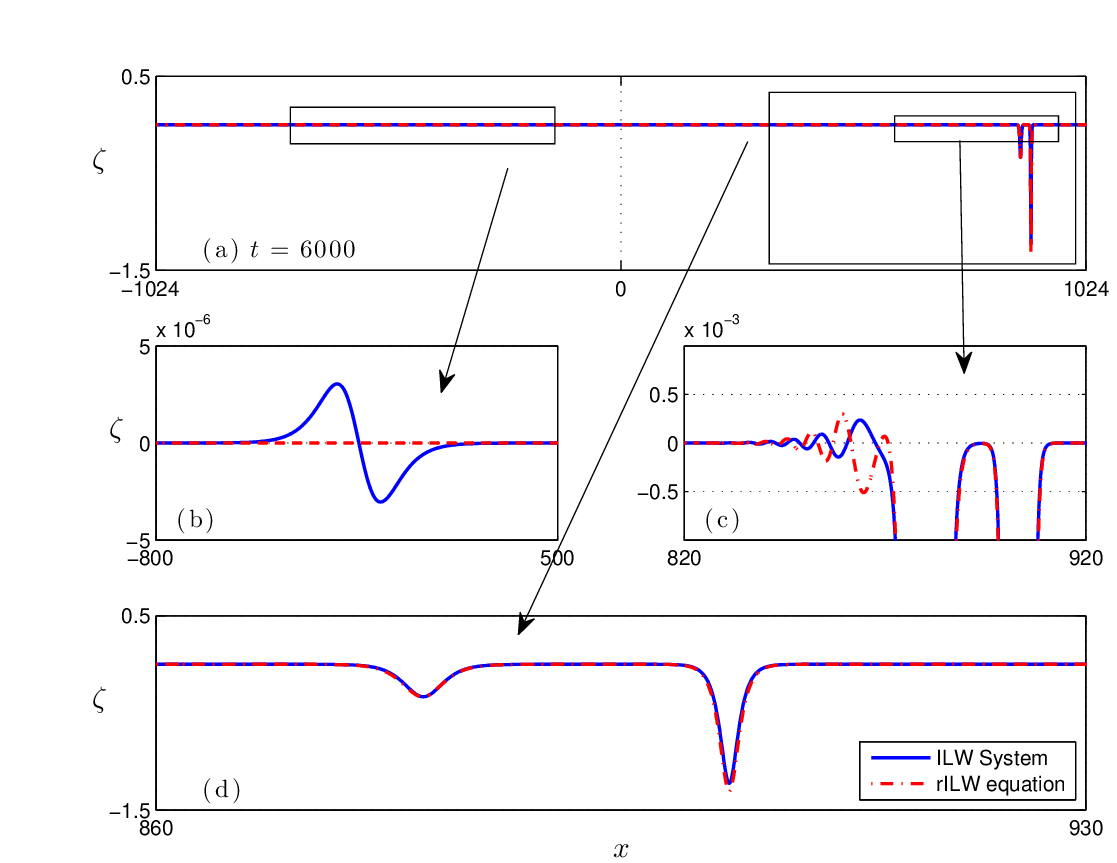}}
\caption{The dispersive tail and the N-shaped wavelet produced after the overtaking collision of two solitary waves of the \acs{ILW} system compared to the solution of the \acs{rILW} equation with the analogous solitary waves, {\it cf.} Figure \ref{fig:figure9}. 
}
\label{fig:figure24}
\end{figure}

The next experiment provides an indication of the similarities between the \acs{BO}- and the \acs{ILW}-systems.   An overtaking collision of two solitary waves with $c_s=0.5$ and $c_s=0.48$ of the \acs{ILW}-system and \acs{rILW}-equation (with $\alpha=1.2$, $\gamma=0.8$, $\varepsilon=\sqrt{\mu}=0.1$ and $\sqrt{\mu_2}=1$) produce similar results. The dispersive tails generated in the case of the \acs{ILW} system consist of the same two parts as in the \acs{BO} system 
and include the N-shaped wavelet.  The collision is thus clearly inelastic, as is the case for the \acs{rILW} equation. The result of these interactions is illustrated in Figure \ref{fig:figure24}.

\subsection{Resolution into solitary waves}\label{sec:resole}

Another important property, which is related to the stability of the solitary waves, is the process whereby solitary waves emerge from the evolution of more general initial conditions. This is often referred to as resolution into solitary waves, and  explains why these special solutions attract so much attention.  To illustrate this resolution property in the current context, a series of runs is reported. The initial conditions in this section are Gaussian perturbations of the interface with zero initial velocity, i.e.
\begin{equation}\label{eq:heap}
\zeta_0(x)=-A e^{-(x/\lambda)^2}~,\quad u_0(x)=0,
\end{equation}
where $A$ and $\lambda$ are constants. Such initial conditions are reminiscent of the early stages of 
tsunami generation \cite{Hammack73}. 
For these initial conditions, the resolution seems to be determined by the energy and mass of the pulse, which in this case is related to the amplitude $A$ and the wavelength $\lambda$, as seen in what follows.

\begin{figure}[!htbp]
\centering
{\includegraphics[width=\columnwidth]{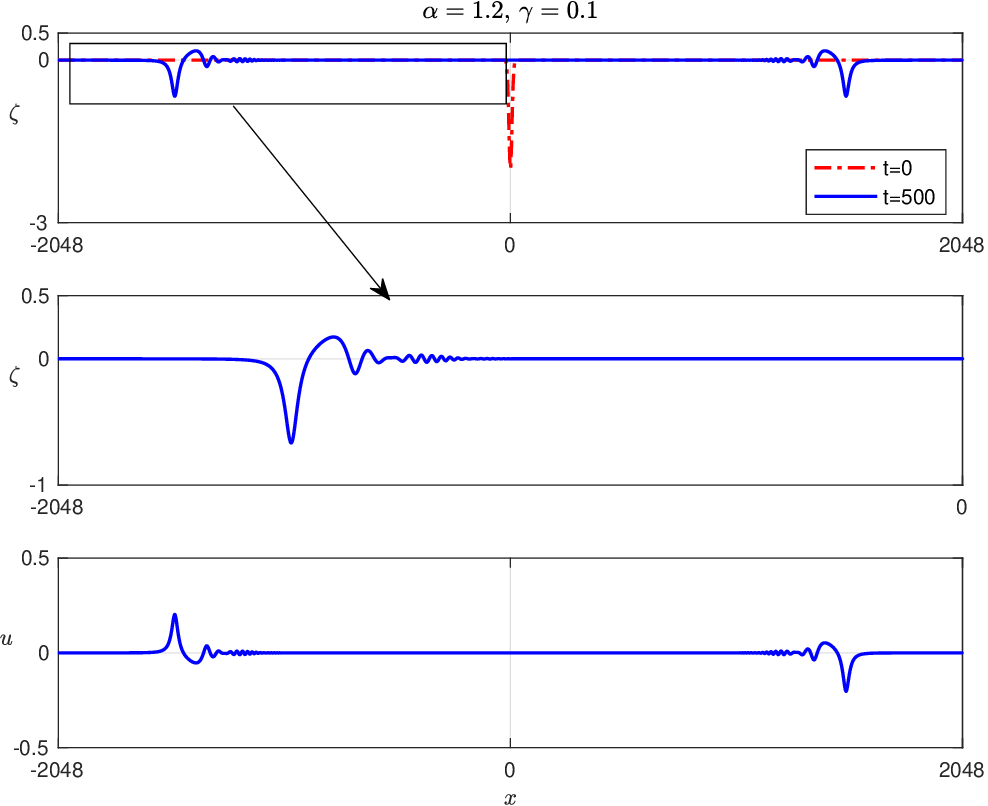}}
\caption{Resolution of a Gaussian initial condition (\ref{eq:heap}) with $A=2$ and $\lambda=10$ into dispersive 
wavetrains for the \acs{BO} system. 
}
\label{fig:figure21}
\end{figure}

Consider the \acs{BO} system with $\alpha=1.2$, $\gamma=0.1$ and $\varepsilon=\sqrt{\mu}=0.1$ with initial conditions (\ref{eq:heap}), where $A=2$ and $\lambda=10$. These initial conditions are unable to trigger the generation of solitary waves. Instead, two counter-propagating dispersive wavetrains appear as shown in Fig. \ref{fig:figure21}.  Indeed, we monitored the $L^\infty$--norm of the solution as a function of $t$ and observed that it decreased steadily as $t$ increased 
from $0$ to $2,000$.  This decay of the maximum norm is one of the hallmarks of a solution containing no 
solitary-wave component.  The result is not surprising, as the branch of solitary-wave solutions of the \acs{BO} 
system appears to have
energies that are 
bounded away from $0$ 

\begin{figure}[!htbp]
\centering
{\includegraphics[width=\columnwidth]{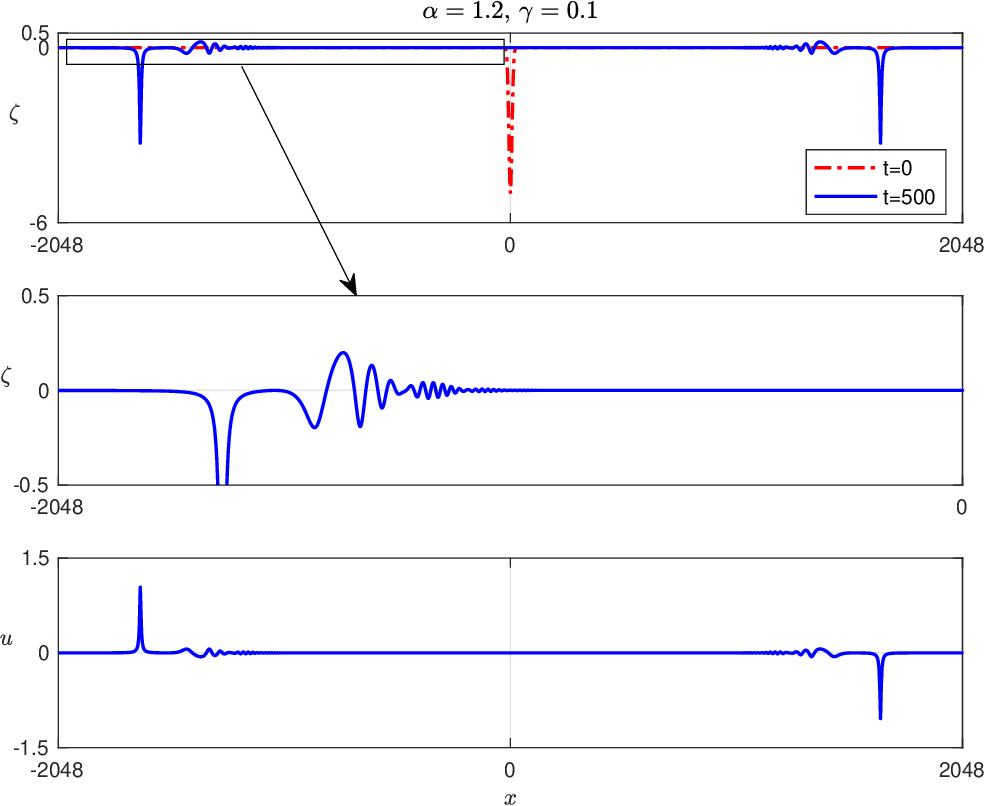}}
\caption{Resolution of a Gaussian initial condition (\ref{eq:heap}) with $A=5$ and $\lambda=10$ into solitary-wave 
solutions of the \acs{BO} system. 
}
\label{fig:figure19}
\end{figure}

\begin{figure}[!htbp]
\centering
{\includegraphics[width=\columnwidth]{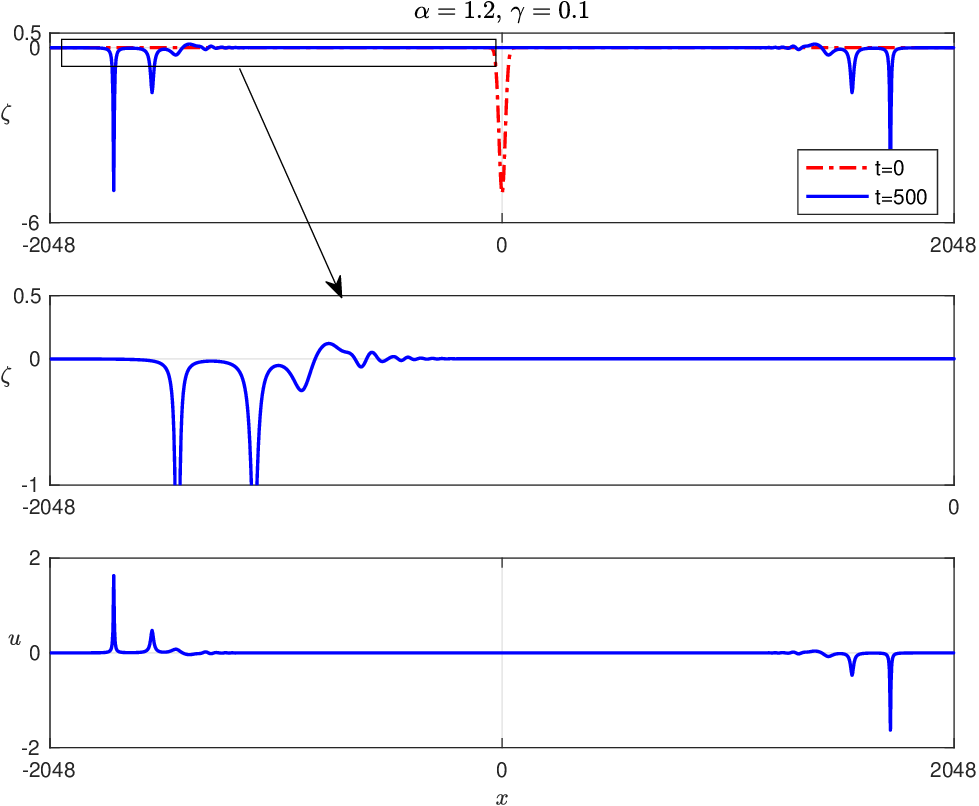}}
\caption{Resolution of a Gaussian initial condition (\ref{eq:heap}) with $A=5$ and $\lambda=20$ into solitary-wave  
solutions of the \acs{BO} system. 
}
\label{fig:figure22}
\end{figure}

\begin{figure}[!htbp]
\centering
{\includegraphics[width=\columnwidth]{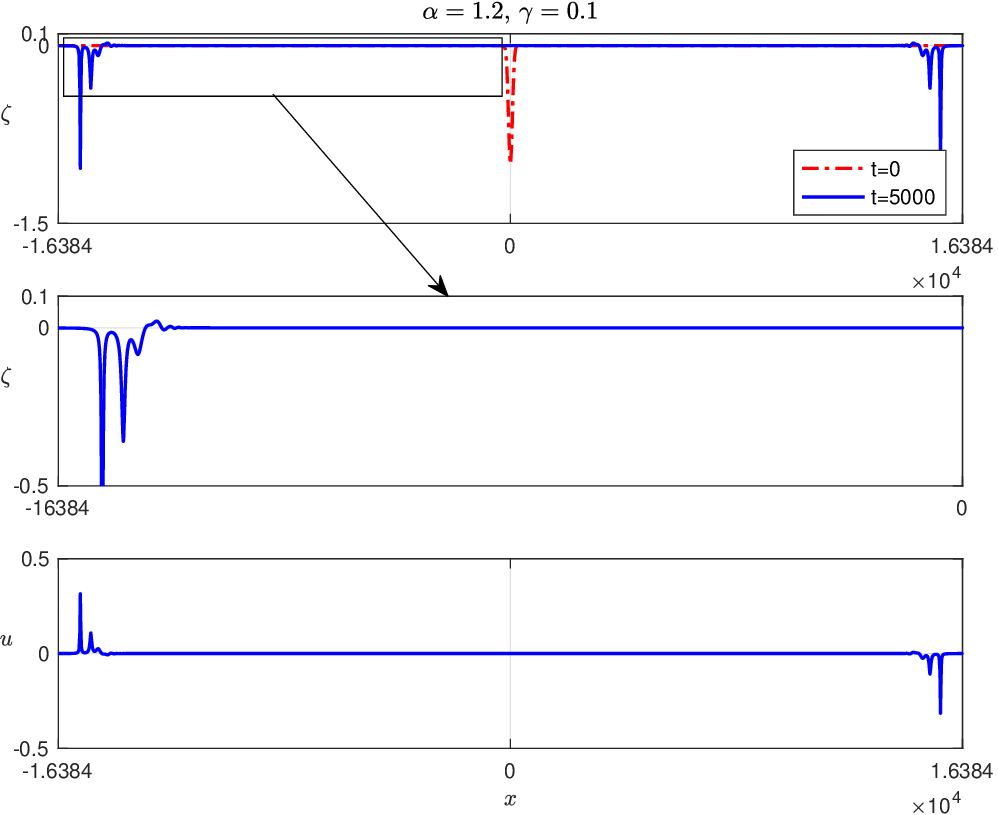}}
\caption{Resolution of a Gaussian initial condition (\ref{eq:heap}) with $A=1$ and $\lambda=100$ into solitary-wave 
solutions of the \acs{BO} system. 
}
\label{fig:figure23}
\end{figure}

Increasing either the amplitude or wavelength of the initial condition has the effect of activating the generation of solitary waves. For example, the evolution of (\ref{eq:heap}) with $A=5$ and $\lambda=10$ is shown in Figure \ref{fig:figure19}. Note that two counter-propagating solitary waves are generated followed by dispersive tails. Similarly, increasing the wavelength of the initial condition by taking $\lambda=20$ and holding $A = 5$ produces the four solitary waves shown in Figure \ref{fig:figure22}, while Figure \ref{fig:figure23} shows the resolution of the initial condition with $A=1$ and $\lambda=100$ into at least four, and very probably six, solitary waves. 

\clearpage

\section{Comparisons with solitary waves of the \acs{BO} equation}\label{sec:dyn}

As discussed in Part I of this work \cite{BDM2021}, the \acs{rBO} and \acs{rILW} equations describe the propagation of internal waves in mainly one direction and have solitary-wave solutions found in closed form. The \acs{BO} and \acs{ILW} systems should be able to describe waves propagating in one direction similarly to the unidirectional models provided that the initial data are `unidirectional'. In the present Section, the ability of the two-way systems to model the waves of their unidirectional counterparts, and especially solitary waves of the \acs{rBO} and \acs{rILW} equations is investigated. 
 The results for the \acs{BO} system are shown below; the experiments for the \acs{ILW} system are easier to perform and the overall conclusions are the same.

 
The tricky point here is that unidirectional models require initial data only for the interfacial variable $\zeta(x,0)=\zeta_0(x)$, whereas the two-way propagation systems also require an initial velocity $u(x,0)=u_0(x)$.  Guided by the rigorous theory 
of surface-wave Boussinesq systems \cite{AABCW}, we use the higher-order approximation 
\begin{equation} \label{high order}
u_0=\sqrt{\gamma(1-\gamma)}\left(\zeta_0+\frac{\varepsilon}{4}\zeta_0^2+\frac{\sqrt{\mu}}{2\gamma}\mathcal{H}\zeta_0\right) 
\end{equation}
to the unknown velocity that applies in the unidirectional case (see  Equation (11) of \cite{BDM2021}).

Thus if we consider a solitary wave $\zeta_0(x)$ of the \acs{rBO} equations, given exactly in Equations (13)--(16) of \cite{BDM2021}, then the unknown initial velocity $u_0(x)$ for the \acs{BO} system is consistently chosen by the higher-order approximation \eqref{high order}.

\begin{figure}[!htbp]
\centering
{\includegraphics[width=\columnwidth]{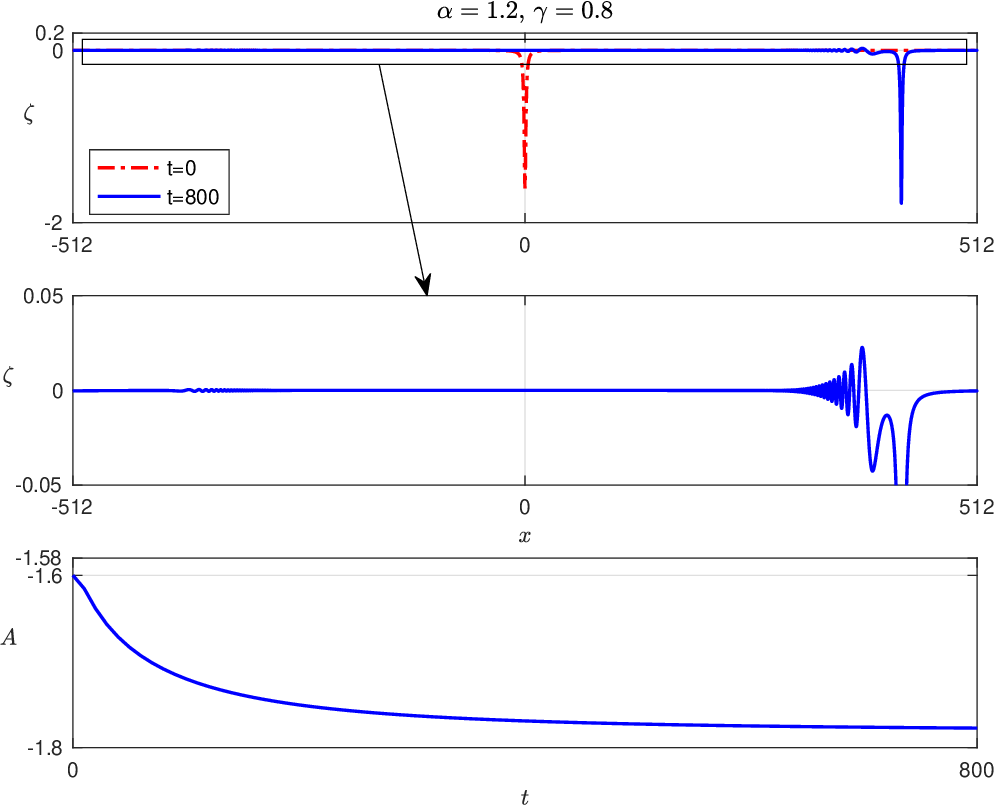}}
\caption{Evolution of a solitary wave of the \acs{rBO} equation used as initial condition for $\zeta_0$ with $u_0$ determined by \eqref{high order} for the \acs{BO} system. The maximum negative excursion of the solution as a function of time is 
also displayed. Observe that the negative excursion levels off as $t$ grows, consistent with a traveling-wave structure.
}\label{fig:oneway}
\end{figure}

For the first reported experiment, we took $\alpha=1.2$, $\gamma=0.8$, $\varepsilon=\sqrt{\mu}=0.1$. Figure \ref{fig:oneway} presents the  evolution of the \acs{BO} system with initial data for $\zeta_0$ taken to be a solitary-wave
solution of the \acs{rBO} equation with $c_s=0.53$ and amplitude $A=-1.6$, together with the $u_0$ derived from \eqref{high order}.   The numerical approximation evolves into a principal profile of solitary-wave type, along with a trailing tail. Most of the resulting wave travels to the right with the \acs{rBO}-solitary wave, while a very small portion travels in the opposite direction. After a simulation which is long enough for the tail to separate and disperse, the numerical approximation appears  to be  a new solitary wave of amplitude $A\approx -1.78$, differing from that of the initial profile by about $11.25\%$ in the maximum norm.  What is particularly notable here is that this `unidirectional' data does indeed 
produce a signal that, in the main, propagates in only one direction.  


\begin{figure}[!htbp]
\centering
{\includegraphics[width=\columnwidth]{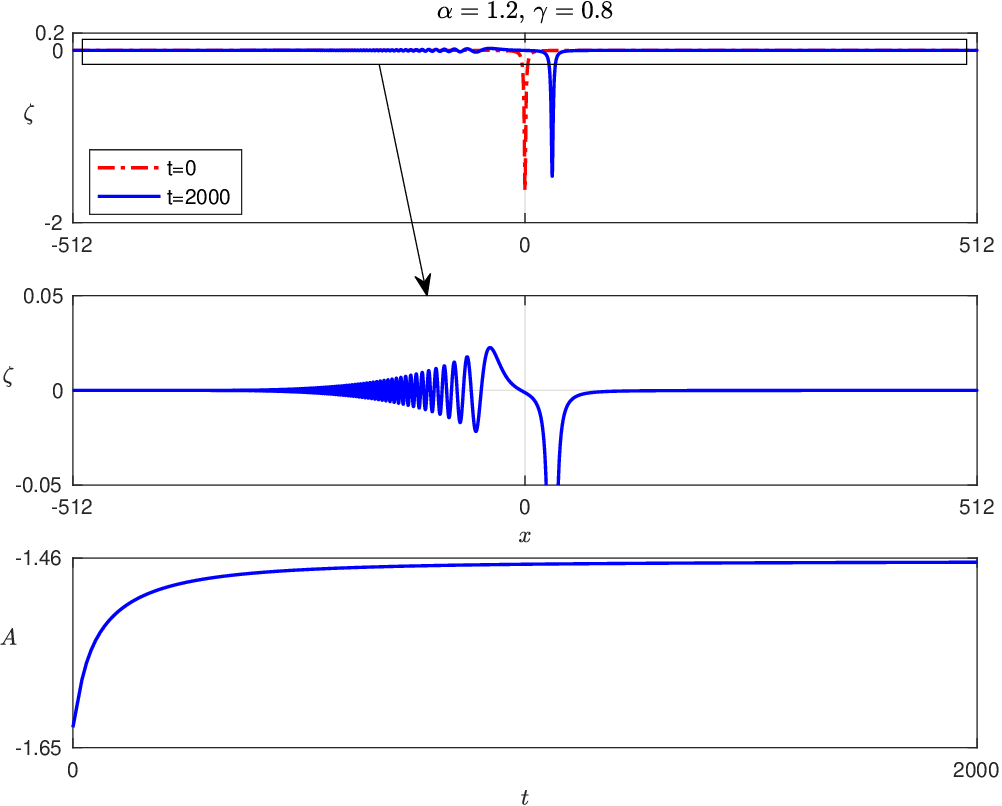}}
\caption{Evolution of a solitary wave of the \acs{BO} system used as an initial condition for the \acs{rBO} equation.  The maximum negative excursion of the solution as a function of time is also shown. 
}\label{fig:oneway2}
\end{figure}

Similarly, a simulation was run of the \acs{rBO} equation with initial condition given by  a solitary-wave solution of the \acs{BO} system. The solitary-wave profile was generated following the numerical techniques introduced in \cite{BDM2021} and with speed $c_s=0.53$. The computed speed-amplitude relation gives a numerical profile of the amplitude $A\approx -1.63$. The IVP for the \acs{rBO} equation is approximated by a fully discrete scheme based on the same procedures described in Section \ref{sec:ummethod} for the \acs{BO} system. The evolution of the corresponding numerical approximation is shown in Figure \ref{fig:oneway2}. As in the previous experiment, this shows  a solution of the \acs{rBO} equation consisting of a solitary-wave profile followed by a tail. As time evolves and the tail disperses, the approximation tends to an emerging solitary wave with amplitude $A\approx -1.47$, which is a difference in uniform norm of about $10\%$ when compared with the amplitude of the initial profile.


\section{Dispersive shock waves}\label{sec:DSWs}

This section is devoted to studying numerically the formation of dispersive shock waves in the regularized \acs{ILW} and \acs{BO} equations, as well as the \acs{ILW} and \acs{BO} systems. Since the results are broadly similar for \acs{ILW} 
and \acs{BO} models, only those corresponding to the \acs{BO}-type models will be shown.

 The formation of \acs{DSW}'s is associated to the propagation of a shock wave through a medium where dispersion effects dominate dissipation (see {\it e.g.} Hoefer and Ablowitz \cite{HoeferA2009}). Under these conditions, the changes in the medium represented by the shock wave typically occur in the form of a structure involving two scales: a rapidly oscillatory wave scale and a slow modulational scale.  Examples of \acs{DSWs} appear in many disciplines such as Water Waves, Plasma Physics, Optics, etc. (see \cite{Whitham1967, Whitham1999,ElH2016} and the references therein).

From the mathematical point of view,  media featuring  \acs{DSW}'s are often described by a system of conservation laws together with dispersive and dissipative terms. In the present study, consideration is given to the situation where dispersion dominates and so dissipation can be safely ignored. For completely integrable equations, the slow modulation of a rapidly oscillating wave can be described using Whitham's modulation theory, \cite{Whitham1967,Whitham1999,Whitham1965,ElH2016}. In such problems, exact solutions for the corresponding Whitham modulation equations can be constructed in terms of the Riemann invariants of the conservation law. These in turn can be used to obtain, from suitable initial and boundary data, asymptotic representations of dispersive shock waves. This was applied, in \cite{GurevichP1974} for the dispersive Riemann problem associated with the KdV equation; see also \cite{Whitham1999,FornbergW1978}.  Other important integrable systems have been similarly analysed; see for example  \cite{Marchant2008} for the mKdV equation, \cite{DobK1991} for the \acs{BO} equation and  \cite{Matsuno1998} and \cite{Lamb1980} for the NLS equation.

In non-integrable systems, it is usually not possible to write the modulation equations in Riemann invariant form and several alternate strategies to study the formation of \acs{DSWs} have  emerged, {\it  e.g.} the method of El \cite{El2005} or the approximate method of Marchant and Smyth discussed in \cite{MarchantS2012}. 

In many cases, \acs{DSW}'s are generated from discontinuities in the initial data. Our computational study will consider two initial-value problems for the \acs{rBO} equation and the \acs{BO} system with discontinuous initial data, namely the Riemann problem and the dam break problem.


Note first that if one neglects the dispersive terms in the \acs{BO} system (\ref{E1}), the hyperbolic  conservation laws
\begin{equation}\label{E31}
\begin{array}{l}
\zeta_t+\frac{1}{\gamma}\left((1-\varepsilon \zeta)u \right)_x=0,\\
u_t+(1-\gamma)\zeta_x-\frac{\varepsilon}{2\gamma}(u^2)_x=0,
\end{array}
\end{equation}
emerge.  This system 
can be written in a matrix-vector form, {\it viz.}
\begin{equation}\label{E32}
{\bf v}+A({\bf v}) {\bf v}_x=0,
\end{equation}
where ${\bf v}=(\zeta, u)^T$ and
$$A=
\begin{pmatrix}-\frac{\varepsilon}{\gamma} u & \frac{1}{\gamma} (1-\varepsilon \zeta)\\
1-\gamma & -\frac{\varepsilon}{\gamma} u
\end{pmatrix}\!.
$$
The distinct eigenvalues of $A$ are
$$\lambda_{\pm}=-\frac{\varepsilon}{\gamma}u \pm \sqrt{\frac{1-\gamma}{\gamma}(1-\varepsilon\zeta)},$$ and the associated eigenvectors, also known  as Riemann invariants of the system (\ref{E32}), are
$$R_{\pm}=\frac{\varepsilon}{\gamma} u\mp 2\ \sqrt{\frac{1-\gamma}{\gamma}(1-\varepsilon\zeta)}.$$ 
They both satisfy the equations
$$\partial_t(R_{\pm})+\lambda_{\pm}\partial_x(R_{\pm})=0.$$
Thus, along the characteristic curves of the system, the quantities $R_{\pm}$ are constant in time.
\subsection{The Riemann problem}
Consider first the \acs{BO} system with Riemann-type initial data, which is to say  the initial conditions are of the form
\begin{equation}\label{E33}
\zeta(x,0)=\left\{
\begin{array}{ll}
\zeta^{-}, & \mbox{ for } x<0\\
\zeta^{+}, & \mbox{ for } x>0
\end{array}
\right. ,
\quad
u(x,0)=\left\{
\begin{array}{ll}
u^{-}, & \mbox{ for } x<0\\
u^{+}, & \mbox{ for } x>0
\end{array}
\right. ,
\end{equation}
with $\zeta^{\pm}, u^{\pm}$ satisfying the compatibility condition
\begin{equation}\label{E34}
\frac{\varepsilon}{\gamma} u^{-}+2\ \sqrt{\frac{1-\gamma}{\gamma}(1-\varepsilon\zeta^{-})}=
\frac{\varepsilon}{\gamma} u^{+}+2\ \sqrt{\frac{1-\gamma}{\gamma}(1-\varepsilon\zeta^{+})}.
\end{equation}
These initial data usually generate what is referred to as a  simple {\acs{DSW}}.  Taking the slightly 
easier case  $\zeta^{+}=u^{+}=0$, 
it transpires that
\begin{equation}\label{E35}
u^{-}=-2\frac{\gamma}{\varepsilon}\left(\sqrt{\frac{1-\gamma}{\gamma}(1-\varepsilon\zeta^{-})} - \sqrt{\frac{1-\gamma}{\gamma}} \right)\! .
\end{equation}

To simulate Riemann-type problems with the numerical method described in Section \ref{sec:ummethod}, the 
initial data must be slightly smoothed.  Otherwise our Fourier-spectral method cannot handle the problem.   Instead of (\ref{E33}), initial data of the form  
\begin{equation}
\label{Ezeta0}
\zeta(x,0)  \ =\   \zeta_0 \left [ 1+\tanh \left(250-|x| \right) \right],
\end{equation}
with $\zeta_0=-1$ and
\begin{equation}
\label{Eu0}
u(x,0)=-2\frac{\gamma}{\varepsilon}\left(\sqrt{\frac{1-\gamma}{\gamma}(1-\varepsilon\zeta(x,0))} - \sqrt{\frac{1-\gamma}{\gamma}} \right)
\end{equation}
is posited.  Both $\zeta_0$ and $u_0$, while smooth, feature large gradients reminiscent of the pure Riemann data 
\eqref{E33} (see Figure 15).

Since $\zeta(-x,0)=\zeta(x,0), \,x\in\mathbb{R}$ and $\zeta(x,0)$ decays to zero exponentially as $|x|\rightarrow\infty$,  the corresponding periodic IVP with initial conditions (\ref{Ezeta0})--(\ref{Eu0}) for the \acs{BO} system and (\ref{Ezeta0}) for the \acs{rBO} equation are integrated  for $x$ in a long spatial interval $(-l,l)$.

\begin{figure}[!htbp]
\centering
{\includegraphics[width=\columnwidth]{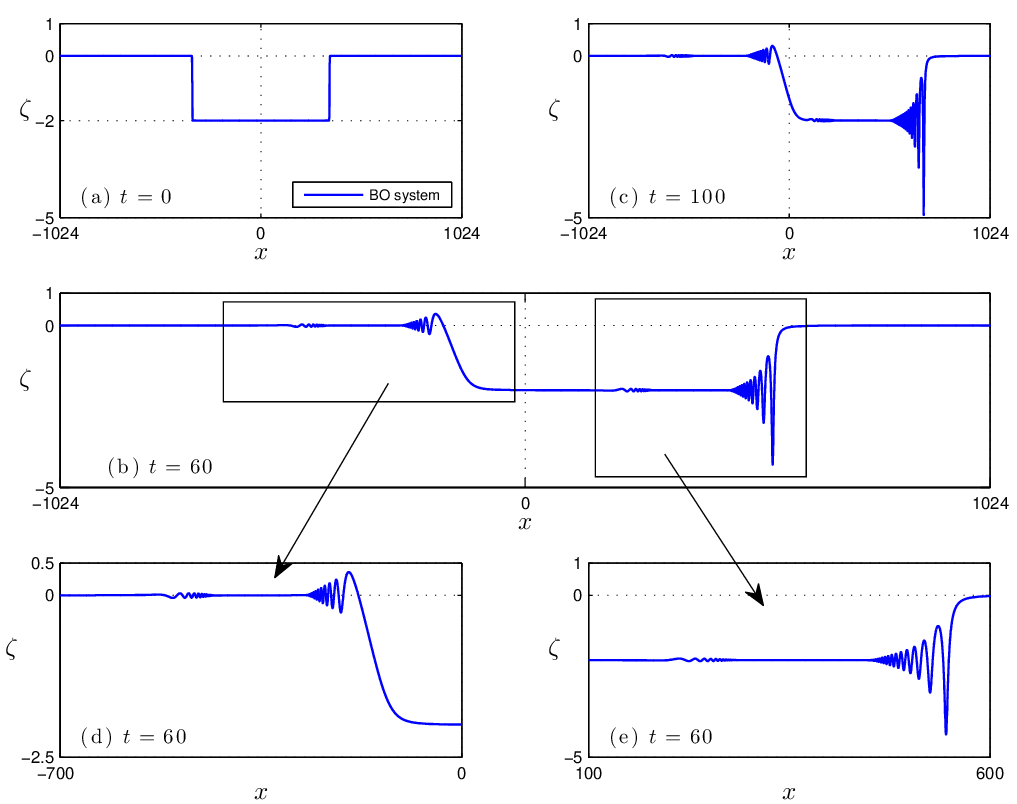}}
\caption{The formation of a simple {\acs{DSW}} for the \acs{BO} system with initial conditions (\ref{Ezeta0})--(\ref{Eu0}) and $\zeta_0=-1$.
}
\label{fig:figure12}
\end{figure}

\begin{figure}[!htbp]
\centering
{\includegraphics[width=\columnwidth]{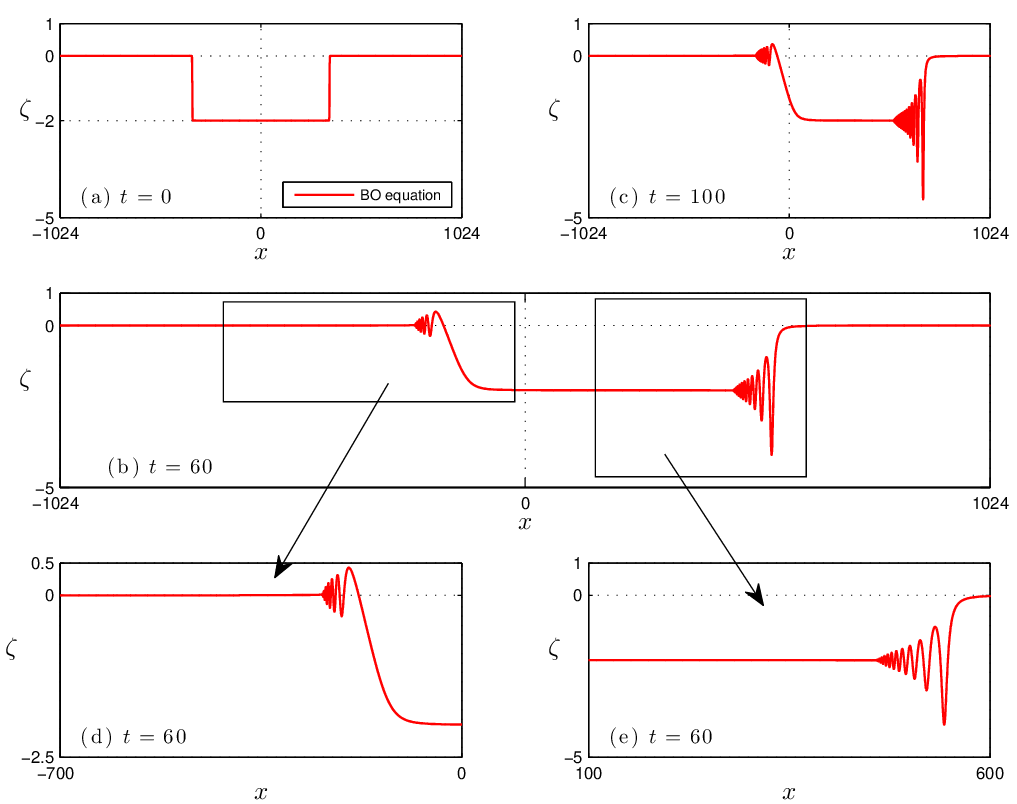}}
\caption{The formation of a simple {\acs{DSW}} for the \acs{rBO} equation and initial condition (\ref{Ezeta0}) with $\zeta_0=-1$.
}
\label{fig:figure13}
\end{figure}

For $l=1024$ and step sizes $h=0.125$ and $\Delta t=0.001$, the numerical simulation with
these initial conditions is shown in Figures \ref{fig:figure12} and \ref{fig:figure13} respectively. 
In the case of the \acs{BO} system, we observe the formation of a right-propagating simple \acs{DSW} plus a dispersive rarefaction wave. Recall that in the absence of the dispersion, one expects a classical shock wave followed by a classical rarefaction wave. Since the compatibility condition (\ref{E34}) is not exact for the \acs{BO} system, the initial conditions generate also leftward propagating dispersive tails. This does not seem to be the case of the \acs{rBO} equation, as shown in Figure \ref{fig:figure13}. Note that the formation of the simple \acs{DSW} and the rarefaction wave does not look to involve additional structures. 


\subsection{The dam break problem}
\begin{figure}[!htbp]
\centering
{\includegraphics[width=\columnwidth]{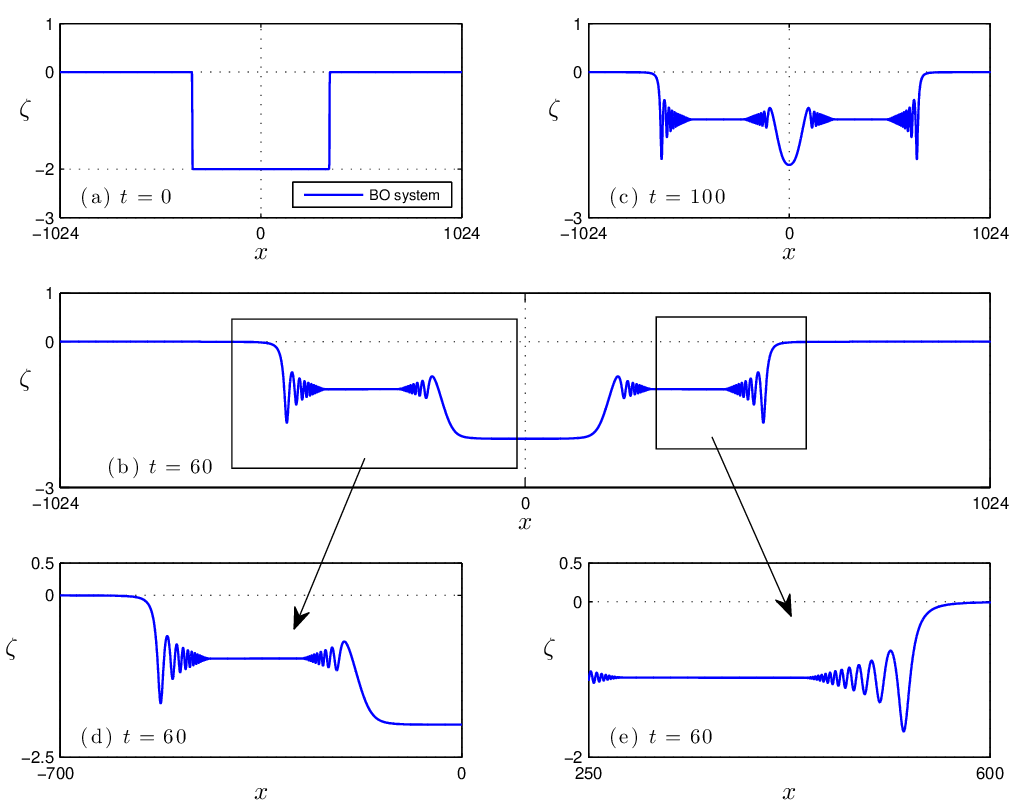}}
\caption{The dam break problem for the \acs{BO} system with initial conditions $\zeta(x,0)$ given by (\ref{Ezeta0}) with $\zeta_0=-1$ and $u(x,0)=0$.
}
\label{fig:figure151}
\end{figure}

Next is introduced the so-called dam break problem. This is a special case of the Riemann problem in which $u^{\pm}=0$ in (\ref{E33}). For the experiment below we will consider the periodic IVP for the \acs{BO} system with initial conditions $\zeta(x,0)$ on $[-1024,1024]$ given by (\ref{Ezeta0}) and $u(x,0)=0$. The resulting numerical simulation, see Figure \ref{fig:figure151}, appears as a waveform consisting of two wave packets, each of which has a dispersive shock plus a rarefaction wave. The packets propagate symmetrically in opposite directions.


\section{Concluding remarks}\label{sec:concluding}
Studied here has been two asymptotic models for the propagation of internal waves along the interface of a two-layer fluid system, with the upper layer bounded above by a rigid lid and the lower layer bounded below by a rigid, featureless horizontal bottom. The systems were derived in \cite{BLS} in the Intermediate Long Wave and the Benjamin-Ono regimes, respectively.  The Benjamin-Ono system is the limiting case of the Intermediate Long-Wave system when a very deep (theoretically infinite) lower layer is assumed. The corresponding one-dimensional pseudo-differential systems (\ref{E1}) have the same quadratic-type nonlinearities, while the linear parts for both of them are non-local.

In Part I of this study \cite{BDM2021}, the authors presented several numerical techniques to generate approximate solitary wave solutions of these systems whose existence, in the small-amplitude case, was proved in \cite{AnguloPS2019}. A comparative study with solitary wave-solutions of related, unidirectional models of \acs{ILW} and \acs{BO} type was also developed.

In the present essay, we further study the solitary-wave solutions of (\ref{E1}), analyzing by computational means some aspects of their dynamics. To this end, the periodic IVP is solved numerically on a long enough spatial interval that 
it approximates well the IVP on $\mathbb R$.  The spatial approximation of our scheme is via a spectral Fourier-Galerkin method (implemented as pseudospectral; the resulting method is algebraically equivalent to a Fourier collocation scheme).  This is coupled with 
an explicit, $4$th-order Runge-Kutta time-stepping. The $L^{2}$--convergence of the spectral semidiscretization was proved in \cite{DDS2021} and the resulting fully discrete scheme was already used in \cite{BDM2021} to check the accuracy of the computed solitary-wave profiles. 
These facts, together with the convergence studies rerported in \cite{BDM2021},  provide confidence in the accuracy of the simulations reported herein.

Specifically, our computational study is concerned with the following aspects: the dynamics of the solitary waves under small and large perturbations, the behaviour of overtaking and head-on collisions of solitary waves, and the resolution of initial data into solitary waves. We also compare computationally the bi-directional and uni-directional models corresponding to the same physical regime.  The study is completed by analyzing numerically the formation of dispersive shock waves. Some of the conclusions of the study are the following:
\begin{itemize}
\item Under small perturbations, the solitary waves seem to be nonlinearly stable, in the sense that an initially perturbed solitary-wave profile evolves into a modified solitary wave with slightly different amplitude and speed, along with a dispersive tail with two components, one traveling to the left and one to the right. The existence of these two dispersive groups was analysed via small-amplitude, plane-wave solutions of (\ref{E1}), linearized about the rest state.
\item Testing the stability of solitary waves of the \acs{BO} and \acs{ILW} systems under large perturbations we found that they are still extremely stable.  They evolve into a new solitary wave of an enhanced magnitude plus a dispersive tail of relatively small amplitude.   
\item The resolution property is illustrated with experiments involving the evolution of initial data of Gaussian type. The initial condition develops into a train of solitary waves whose number depends on the {\it energy} of the initial Gaussian pulse (represented through its amplitude and wavelength parameters) and with a dispersive tail behind.
\item The interactions are observed to be inelastic, something that is expected on account of the lack of Hamiltonian structures and the fact that there are only linear conserved quantities. Since the models are bidirectional, two types of collisions are considered. In the case of overtaking collisions, the behaviour of the waves after the interaction was compared to that of similar experiments for the uni-directional regularized \acs{BO} equation. The main difference observed in the experiments seems to be the formation, in the case of the \acs{BO} system, of $N$-shaped wavelets in the tail, traveling in the direction opposite to that of the emerging solitary waves. In the case of head-on collisions, the interactions develop dispersive tails propagating in both directions, as expected from the analysis of the small solutions of the linearized system, mentioned above.
\item Comparisons between the unidirectional and bidirectional models when the initial data for the bidirectional model 
is well prepared for unidirectionality show remarkable similarities.  This suggests that a theorem of the sort derived in \cite{AABCW} for surface waves likely holds for internal waves as well.  
\item The numerical experiments concerning the formation of \acs{DSW}s were focused on the classical Riemann and dam break problems. In the case of the Riemann problem, suitable initial data generated a simple \acs{DSW} along with a dispersive rarefaction wave, with an additional small dispersive structure traveling in the opposite direction. We checked that from the same initial condition, the \acs{BO} and regularized \acs{BO} equations developed similar structures but without the small additional dispersion. In the case of the dam break problem, the experiments suggest that from initial data with zero flow, the approximate solution of the \acs{BO} system develops two \acs{DSWs} plus rarefaction structures traveling in opposite directions.
\end{itemize}

\section*{Acknowledgments}
JB acknowledges support from the University of Illinois at Chicago College of Arts and Science.
AD is supported by the Spanish Agencia Estatal de Investigaci\'on under
Research Grant  PID2020-113554GB-100, by the Junta de Castilla y Le\'on and FEDER funds (EU) under Research Grant
VA193P20.

\section*{Conflict of interest}
On behalf of all authors, the corresponding author states that there is no conflict of interest.

\section*{Data availability statement}
The datasets generated during and/or analysed during the current study are available from the corresponding author on reasonable request.



\bibliographystyle{elsarticle-num}

\begin{thebibliography}{10}
\expandafter\ifx\csname url\endcsname\relax
  \def\url#1{\texttt{#1}}\fi
\expandafter\ifx\csname urlprefix\endcsname\relax\def\urlprefix{URL }\fi
\expandafter\ifx\csname href\endcsname\relax
  \def\href#1#2{#2} \def\path#1{#1}\fi

\bibitem{BDM2021}
J.~L. Bona, A.~Dur\'an, D.~Mitsotakis, {Solitary-wave solutions of Benjamin-Ono
  and other systems for internal waves. I. Approximations}, Discrete Cont.
  Dynamical Sys. B {\bf 41} (2021) 87--111.
\newblock \href {https://doi.org/10.3934/dcds.2020215}
  {\path{doi:10.3934/dcds.2020215}}.

\bibitem{BLS}
J.~L. Bona, D.~Lannes, J.-C. Saut, Asymptotic models for internal waves, J.
  Math. Pures. Appl. {\bf 89} (2008) 538--566.

\bibitem{Xu2012}
L.~Xu, {Intermediate long wave systems for internal waves}, Nonlinearity {\bf
  25} (2012) 597--640.

\bibitem{CGK2005}
W.~Craig, P.~Guyenne, H.~Kalisch, Hamiltonian long-wave expansions for free
  surfaces and interfaces, Comm. Pure Appl. Math. {\bf 58} (2005) 1587--1641.

\bibitem{AnguloPS2019}
J.~Angulo-Pava, J.-C. Saut, Existence of solitary-wave solutions for internal
  waves in two-layer systems, Quart. Appl. Math., 78 (2020) 75--105.

\bibitem{DDS2021}
V.~A. Dougalis, A.~Dur\'an, L.~Saridaki, Numerical solution of internal-wave
  systems in the {I}ntermediate {L}ong {W}ave and the {B}enjamin-{O}no regimes,
  Bull. Hellenic Math. Soc. {\bf 66} (2022) 11--25.

\bibitem{HoeferA2009}
M.~A. Hoefer, M.~J. Ablowitz, Dispersive shock waves, Scholarpedia 4 (2009)
  5562.

\bibitem{ElH2016}
G.~A. El, M.~A. Hoefer, {Dispersive shock waves and modulational theory},
  Physica D 333 (2016) 11--65.

\bibitem{zabusky}
N.~J. Zabusky, M.~D. Kruskal, Interaction of ``solitons" in a collisionless
  plasma and the recurrence of initial states, Phys. Rev. Lett. 15 (1965)
  240--243.

\bibitem{tappert}
F.~Tappert, Numerical solution of the {K}orteweg-de {V}ries equation and its
  generalization by the split-step {F}ourier method, Lectures in Applied
  Mathematics 15 (Ameirican Math. Soc., Providence, 1974) 215--216.

\bibitem{bona}
J.~L. Bona, Convergence of periodic wave trains in the limit of large
  wavelength, Appl. Sci. Res. {\bf 37} (1981) 21--30.

\bibitem{hchen}
H.~Chen, {Long-period limit of nonlinear, dispersive waves: The{ BBM}
  equation}, Differential Integral Eq. {\bf 19} (2006) 463--480.

\bibitem{CHQZ}
C.~Canuto, M.~Y. Hussaini, A.~Quarteroni, A.~T. Zang, {Spectral Methods in
  Fluid Dynamics}, Springer; New York, 1985.

\bibitem{DDMM}
V.~A. Dougalis, A.~Dur\'an, M.~A. Lopez-Marcos, D.~Mitsotakis, A numerical
  study of the stability of solitary waves of the {B}ona-{S}mith family of
  {B}oussinesq systems, J. Nonlinear Sci. {\bf 17} (2007) 569--607.

\bibitem{KB2000}
H.~Kalisch, J.~L. Bona, Models for internal waves in deep water, Discrete
  Contin. Dynamical Syst. {\bf 6} (2000) 1--20.

\bibitem{K2005}
H.~Kalisch, {Error analysis of spectral projections of the regularized
  Benjamin-Ono equation}, BIT {\bf 45} (2005) 69--89.

\bibitem{DM2008}
V.~A. Dougalis, D.~E. Mitsotakis, {Theory and numerical analysis of Boussinesq
  systems: A review}, in: N.~A. Kampanis, V.~A. Dougalis, J.~A. Ekaterinaris
  (Eds.), Effective Computational Methods in Wave Propagation, CRC {P}ress;
  Boca Raton, 2008, pp. 63--110.

\bibitem{DKM2011}
D.~Dutykh, T.~Katsaounis, D.~Mitsotakis, {Finite volume schemes for dispersive
  wave propagation and runup}, J. Comp. Phys. {\bf 230} (2011) 3035--3061.

\bibitem{DKM2012}
D.~Dutykh, T.~Katsaounis, D.~Mitsotakis, {Finite volume methods for
  unidirectional dispersive wave models}, Int. J. Num. Meth. Fluids {\bf 71}
  (2013) 717--736.

\bibitem{Hammack73}
J.~Hammack, A note on tsunamis: their generation and propagation in an ocean of
  uniform depth, J. Fluid Mech. 60~(4) (1973) 769--799.

\bibitem{AABCW}
A.~A. Alazman, J.~P. Albert, J.~L. Bona, M.~Chen, J.~Wu, Comparisons between
  the {BBM} equation and a {B}oussinesq system, Advances Differential Eq. {\bf
  11} (2006) 121--166.

\bibitem{Whitham1967}
G.~B. Whitham, {Non-linear dispersion of water waves}, J. Fluid Mech. 27 (1967)
  399--412.

\bibitem{Whitham1999}
G.~B. Whitham, {Linear and Nonlinear Waves}, John Wiley \& Sons Inc.; Hoboken,
  New Jersey, 1999.

\bibitem{Whitham1965}
G.~B. Whitham, {A general approach to linear and non-linear dispersive waves
  using a Lagrangian}, J. Fluid Mech. {\bf 22} (1965) 273--283.

\bibitem{GurevichP1974}
A.~V. Gurevich, L.~P. Pitaevskii, Nonstationary structure of a collisionless
  shock wave, Sov. Phys. JETP 38 (1974) 291--297.

\bibitem{FornbergW1978}
B.~Fornberg, G.~B. Whitham, A numerical and theoretical study of certain
  nonlinear wave phenomena, Phil. Trans. R. Soc. Lond. A 289 (1978) 373--404.

\bibitem{Marchant2008}
T.~R. Marchant, Undular bores and the initial-boundary value problem for the
  modified {K}orteweg-de {V}ries equation, Wave Motion 45 (2008) 540--555.

\bibitem{DobK1991}
S.~T. Dobrokhotov, I.~M. Krichever, Multiphase solutions of the
  {B}enjamin-{O}no equation and their averaging, Matematicheskie Zametki 49
  (1991) 42--50.

\bibitem{Matsuno1998}
Y.~Matsuno, {Nonlinear modulation of periodic waves in the small dispersion
  limit of the Benjamin-Ono equation}, Phys. Rev. E {\bf 58} (1998) 7934--7940.

\bibitem{Lamb1980}
G.~L. Lamb, {Elements of Soliton Theory}, John Wiley \& Sons Inc., 1980.

\bibitem{El2005}
G.~A. El, {Resolution of a shock in hyperbolic systems modified by weak
  dispersion}, Chaos 15 (2005) 037103.

\bibitem{MarchantS2012}
T.~R. Marchant, N.~F. Smyth, Approximate techniques for dispersive shock waves
  in nonlinear media, J. Non. Opt. Phys. and Mat. 21 (2012) 1250035.

\end{thebibliography}

\end{document}